\documentclass{aa}

\usepackage[varg]{txfonts}
\usepackage{lineno}
\usepackage{bbold}
\usepackage{bm}
\usepackage{graphicx}
\usepackage{float}
\usepackage{amssymb}
\usepackage{amsfonts}
\usepackage{amsmath}
\usepackage{color}
\usepackage{natbib}
\bibpunct{(}{)}{;}{a}{}{,}
\usepackage{hyperref}
\usepackage{longtable}
\usepackage{pdflscape}
\usepackage{dirtytalk}
\usepackage{multirow}
\usepackage{hhline}
\usepackage{soul}
\usepackage{booktabs}

\usepackage{mathtools}

\newcommand{\lensname}{WFI2033--4723}

\usepackage{physics}

\newcommand{\e}[1]{\texttt{#1}}

\usepackage{xcolor}
\usepackage{enumitem}

\setlist[enumerate,1]{label=\arabic*., left=0.25cm}

\usepackage[nameinlink]{cleveref}
\usepackage{cleveref}
\crefname{section}{Sect.}{Sects.}
\Crefname{section}{Sect.}{Sects.}

\crefname{subsection}{Sect.}{Sects.}
\Crefname{subsection}{Sect.}{Sects.}

\crefname{subsubsection}{Sect.}{Sects.}
\Crefname{subsubsection}{Sect.}{Sects.}

\crefname{figure}{Fig.}{Figs.}
\Crefname{figure}{Fig.}{Figs.}

\crefname{table}{Table}{Tables}
\Crefname{table}{Table}{Tables}

\crefname{equation}{Eq.}{Eqs.}
\Crefname{equation}{Eq.}{Eqs.}

\crefname{appendix}{Appendix}{Appendices}
\Crefname{appendix}{Appendix}{Appendices}

\makeatletter
\renewcommand\@seccntformat[1]{\csname the#1\endcsname\quad}
\makeatother
\usepackage{tabularx}
\usepackage{arydshln}
\usepackage[thinc]{esdiff}

\usepackage{siunitx}
\usepackage{textcomp}

\usepackage{threeparttable}

\newcommand{\kms}{\si{\kilo\meter\per\second}}
\let\oldAA\AA
\renewcommand{\AA}{\textup{\oldAA}}

\newcommand{\lowup}[2]{\raisebox{0.5ex}{\tiny$^{+#2}_{-#1}$}}

\usepackage{tablefootnote}

\usepackage{placeins}
\usepackage{dblfloatfix}
\usepackage{cuted}
\usepackage{capt-of}
	
\begin{document}

	\title{TDCOSMO: XX. WFI2033--4723, the first quadruply imaged quasar modeled with JWST imaging}
	
	\author{
		D. M. Williams\thanks{e-mail: devon@astro.ucla.edu}\fnmsep\inst{1}
		\and
		T. Treu\inst{1}
		\and
		S. Birrer\inst{2}
		\and
		A. J. Shajib\fnmsep\thanks{NHFP Einstein Fellow}\fnmsep\inst{3, 4, 5}
		\and
		K. C. Wong\inst{6}
		\and
		T. Morishita\inst{7}
            \and
		T. Schmidt\inst{1}
		\and
		M. Stiavelli\inst{8}
	}
	
	\institute{
		Department of Physics and Astronomy, UCLA, Los Angeles, CA 90095, USA
		\and
		Department of Physics and Astronomy, Stony Brook University, Stony Brook, NY 11794, USA
		\and
		Department of Astronomy and Astrophysics, University of Chicago, Chicago, IL 60637, USA
        \and
        Kavli Institute for Cosmological Physics, University of Chicago, Chicago, IL 60637, USA
        \and
        Center for Astronomy, Space Science and Astrophysics, Independent University, Bangladesh, Dhaka 1229, Bangladesh
		\and
		Research Center for the Early Universe, Graduate School of Science, The University of Tokyo, Tokyo 113-0033, Japan
		\and
		IPAC, California Institute of Technology, Pasadena, CA 91125, USA
            \and
		Space Telescope Science Institute, Baltimore, MD 21218, USA
	}

	\date{Received 00 Month 0000; accepted 00 Month 0000}
	
	\abstract{
		Gravitational time delays offer unique, independent measurements of the Hubble constant, $H_0$. Precise measurements of $H_0$ stand as one of the most pressing challenges in modern cosmology, and obtaining them with time delays requires precise lens models. While much work has focused on streamlining the modeling process to keep pace with the erumpent discovery of strongly lensed systems, a critical step toward reducing uncertainty in $H_0$ involves increasing the precision of individual lens models themselves. In this work, we demonstrate that the unprecedented imaging capabilities of JWST make this goal attainable. We present the first lens model for time-delay cosmography derived from JWST data, applied to the quadruply imaged quasar WFI2033--4723. While the primary source of systematic uncertainty in time-delay cosmography is currently the mass-sheet degeneracy (MSD), the sensitivity of models to this MSD varies depending how the point spread function (PSF) errors are mitigated. As the PSF is also the primary source of uncertainty in lens modeling, we focus on a comparison of different PSF modeling methods. Within the context of power-law models, we recover results in agreement with previous Hubble Space Telescope (HST)-based models, but with better precision of key lensing parameters through the implementation of new PSF modeling techniques. Despite the record-holding precision of this system's HST modeling, we were able to achieve an additional $22\%$ increase in precision of the Fermat potential difference, thus directly reducing uncertainties of cosmological inference. These results would produce a 3\% ($1\sigma$ of the lens modeling error) shift of H$_0$ toward a higher value for this lens, if one were to keep all else constant.
        This work substantiates the groundbreaking potential of JWST for time-delay cosmography and lays the groundwork for modeling systems previously too faint to provide meaningful constraints on $H_0$.}
	
	\keywords{
	Gravitational lensing: strong - 
	Methods: data analysis, statistical - 
        Galaxies: active, distances and redshifts - 
        cosmological parameters - 
        distance scale
	}
	
	\maketitle

        \titlerunning{TDCOSMO: XX. WFI2033--4723}
        \authorrunning{D. M. Williams et al.}

\section{Introduction}
\label{sec:intro}

The expansion rate of the Universe remains one of the most hotly debated topics in cosmology, even after two decades of measurements with improving statistical precision. Measurements of the early Universe, extrapolating to what the current value of the expansion rate should be given today's standard cosmological model \text{$\Lambda \text{CDM}$}, give values in tension with direct measurements of the late Universe. Hope that the tension would dissolve in time is proving to be misplaced: As the precision of both the early- and late-Universe measurements have increased, so too has their tension. (For a comprehensive review, see \citet{verde2019,shah2021,abdalla2022}.)

The confirmation of the ``Hubble tension" would have profound implications for cosmology, requiring adjustments to or even a complete revision of the standard model \citep[e.g.,]{knox2020, efstathiou2021}. Considerable effort has been spent evaluating sources of systematic uncertainties in both methods, with none proving successful \citep{riess2019,  riess2021, riess2022, freedman2019, freedman2020, dainotti2021, mortsell2022}. It is clear that multiple independent methods with comparable precision will be needed to achieve a final verdict on residual systematics error and settle the Hubble tension once and for all.

One such method is time-delay cosmography, which was first discussed in \cite{refsdal1964} \citep[for recent reviews, see, e.g.,][]{treu2016,treu2022}. It provides a one-step measurement of $H_0$ independent of methods that rely on the local distance ladder for late-Universe measurements or on sound horizon physics, such as Cosmic Microwave Background (CMB) measurements, for early-Universe measurements \citep[e.g.,][]{vanderriest1989, keeton1997, oguri2007, suyu2010b}. In this work, we focus on time-delay cosmography applied to lensed quasars. Quadruply imaged quasars (hereafter quads), where light from a quasar is bent around a lensing galaxy to produce four distinct images, provide three independent measurements of the expansion rate along with tight constraints on the lensing galaxy's mass distribution.

Until the past decade, the precision of time-delay cosmography was limited by the small number of known systems and further by the number of those systems with reliable time delays measured. Fortunately, there has been an explosion of newly discovered strong lenses, resulting in a commensurate effort to gather the necessary ancillary data and to quicken the lens modeling process. Key advances include: systematic time-delay measurements (spearheaded by the COSMOGRAIL project, e.g., \citealt{millon2020, bonvin2019}), automated modeling techniques (e.g., \citealt{shajib2019, schmidt2022}), consistent analyses of lens environments (e.g., \citealt{wells2024a}), and improved kinematic constraints (e.g., \citealt{tan2024}). Unfortunately, ground-based seeing-limited images cannot be modeled to the precision required for cosmography. The Hubble Space Telescope (HST), with its stable point spread function (PSF), has been the main workhorse for these kinds of models \citep[e.g.,][]{suyu2010,birrer2019a}, even though some breakthroughs have been achieved in the use of adaptive optics assisted data \cite[e.g.,][]{chen2019}.

In this context, JWST plays multiple important roles: First, it holds the potential to further improve the precision of lens models; second, it provides an entirely new means of verifying systematics in previous models that are based on other telescope data (namely HST); third, with its superior resolution and sensitivity, it will be crucial to exploit the large number of systems about to be discovered, which will likely be fainter than the ones analyzed so far \citep{oguri2010}; and fourth, the extended wavelength range will enhance the detection of the extended emission (arcs) from the quasar host galaxy, providing tighter constraints on the deflector's mass distribution.

 \begin{figure}[ht]
    \centering
    \includegraphics[width=0.5\textwidth]{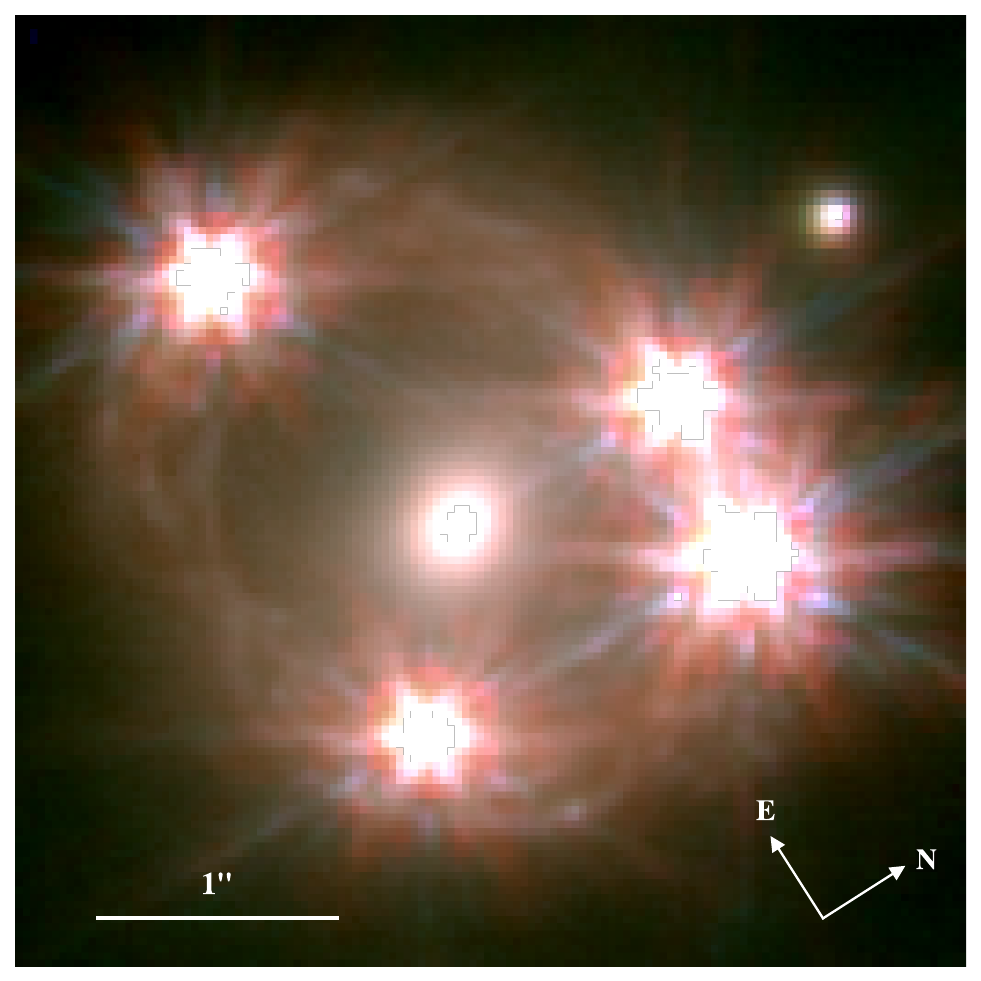}
    \caption{JWST NIRCam color image of WFI2033-4723 (GTO Program \#1198; PI: Stiavelli). The F115W band is mapped to the luminosity and blue, F150W to cyan, F277W to yellow, and the F356W band to red. Near the bottom image, there is a bright spot on the lower-right side that maps back to other bluer regions of the ring, indicating it is a multiply imaged star-forming region contained in the z$\sim\!1.6$ host. Preliminary models place the physical size at $\sim\!150$pc and locate it at $\sim\!2$kpc from the center of the galaxy (assuming a flat CDM cosmology with h = 0.7 and m = 0.3).}
    \label{fig:wfi2033_color_image}
\end{figure}

In this paper, we present the first JWST-based quadruply imaged quasar lens model for cosmology. We focus on the lens system \lensname  discovered by \cite{morgan2004} and previously modeled based on HST data \citep{rusu2020}. This system is particularly interesting, as the lensed host galaxy is detected by both HST and JWST but only clearly resolved by JWST, as shown in  \Cref{fig:hst_jwst_comp}. The figure also illustrates the complexity of JWST's PSF, which represents a major challenge of this work. Further, the system is fairly complex, having a satellite and multiple nearby galaxy perturbers with a non-negligible impact on the lens models (and so require explicit modeling) \citep{sluse2019,rusu2020}. The goals of this paper are thus the following:
\begin{enumerate}
    \item To verify whether the HST and JWST models agree for this system, justifying our handling of systematic effects in the HST models.
    \item To quantify the improvements in precision that JWST provides for lens modeling.
    \item To demonstrate sufficient handling of the PSF to facilitate these improvements.
    \item To determine if the superior angular resolution and improved extended-emission (or arc) detection of JWST yields improved constraints on nearby perturbers.
\end{enumerate}

As the goal of this paper is to make a direct comparison of strong lens modeling based on HST and JWST, we limit ourselves to the core analysis and do not perform a full determination of the Hubble constant based on this study. While this work can be used for such work, a full analysis will be performed in a proceeding paper. Among the simplifications, we work within the context of a power-law model and do not consider the mass sheet-degeneracy (MSD). We also do not consider updates to the external convergence estimates that have been implemented by our collaboration since the \citet{rusu2020} analysis. These choices make it more straightforward to perform a direct comparison with \citet{rusu2020}. For these reasons, we compare our results in terms of Fermat potential differences between the images with the longest expected time delay $\Delta \tau_{\rm BC}$ (for brevity, we refer to this quantity as "the Fermat potential" in the text). This is the quantity more directly connected to the lens model yet close enough to $H_0$ to permit an intuitive understanding of our results. All other things being equal, this quantity is directly proportional to $H_0$. However, we stress that all other things are not equal, since other factors contribute to $H_0$, for example, the external convergence, determination of the MSD parameter from the velocity dispersion, and the other time delays of the quad. Therefore, our result should not be taken out of context and plugged into the analysis of \citet{rusu2020} directly to update their measurement of $H_0$.

An important difference with respect to the previous analysis is that we use a different strong lensing code: \citet{rusu2020} used \textsc{glee} \citep{suyu2010b}, while we use \textsc{lenstronomy} \citep{birrer2018, birrer2021}. The two codes have been tested extensively and compared directly \cite[e.g.,][]{shajib2022}. The comparison showed that they give consistent results even though they differ in the way they parametrize important components, such as the host galaxy surface brightness distribution. When possible, we use the same parametrization as \citet{rusu2020} to facilitate comparison.

The main result of this paper is that JWST’s PSF can be modeled with sufficient accuracy to enable the development of the next generation of lens models for time-delay cosmography. We show that JWST yields better precision than HST and that the results agree within the statistical errors. We also show that with JWST's superior resolution and sensitivity, the mass distribution of the satellite galaxy is better constrained than with HST.

The paper is organized as follows. First, in \Cref{sec:ana} we outline the analysis steps. Then in \Cref{sec:data} we present the data. In \Cref{sec:psf_modeling} we describe our PSF reconstruction methodology. In \Cref{sec:mod} we describe our modeling choices. In \Cref{sec:comb} we discuss how our models are combined in the final inference. In \Cref{sec:results_and_discussion} we present our results and compare them with the previous HST model. In \Cref{sec:conclusion} we provide a brief summary.

When necessary we adopt a standard flat $\Lambda$CDM cosmology with $H_0$ = 70 \kms, $\Omega_{\rm,m}$=0.3, and $\Omega_{
\rm b}=0.05$. However, we note that this choice is irrelevant from the point of view of the comparison presented here since lens modeling is done in angular coordinates.

\section{Outline of analysis}
\label{sec:ana}

The big-picture goal of this paper is to assess how advancements in space-based imaging quality impact the precision and accuracy of model-derived parameters. Specifically, we investigate how the transition from HST to JWST imaging influences the Fermat potential differences, $\Delta \tau$, which are crucial outputs for cosmographical inference of $H_0$. To this end, we first review the lensing theory and discuss the observables in \Cref{ssec:ana_theory}, we then review inherent degeneracies and systematic concerns of the analysis in \Cref{ssec:msd} and how we address them in \Cref{ssec:ana_kinematics}, and we finish with the outline of our Bayesian analysis in \Cref{ssec:ana_bayes}.

\subsection{Theory and observables }
\label{ssec:ana_theory}

This section sprints through the theoretical groundwork for time-delay cosmography. For more of a ``stroll,'' we refer to recent reviews, such as \cite{treu2016} or \cite{suyu2018}.

As light is deflected by gravitational fields, multiple rays of light can occasionally be observed traversing different paths even though they originate from the same object. This special class of objects that show this phenomena are described as ``multiply imaged." The combination of these two factors, traveling different paths despite starting from the same position, allows one to calculate distances at cosmological scales.

How an image is distorted is specified by the lens equation
\begin{equation}
	\bm{\beta} = \bm{\theta} - \bm{\alpha} (\bm{\theta}),
\end{equation}
with the lensed position in the lens plane ($\bm{\theta} = (\theta_1, \theta_2)$), the position in the source plane ($\bm{\beta} = (\beta_1, \beta_2)$), and the deflection angle ($\bm{\alpha}$).
If we assume the deflection occurs in a single lens plane, we can quantify the observed delay between images with the Fermat potential:
\begin{equation}\label{eq:tau}
\tau (\bm{\theta}, \bm{\beta}) \equiv \left[ \frac{(\bm{\theta} - \bm{\beta})^2}{2} - \psi(\bm{\theta}) \right],
\end{equation}
with $\psi(\bm{\theta})$ as the lens potential. The Fermat potential is, up to an affine transformation, the light travel time along a ray starting at $\bm{\beta}$, passing through $\bm{\theta}$ in the lens plane, and arriving at the observer. The time required for light to reach the observer from one of these images is influenced by two factors. First is the spatial path length the light must traverse, and second is the gravitational potential the light experiences along that path \citep[via the `Shapiro delay'][]{shapiro1964}.

The lens potential ($\psi(\bm{\theta})$) is defined such that it solves two conditions: first, for the deflection angle,
\begin{equation}
	\bm{\alpha} (\bm{\theta}) = \nabla \psi (\bm{\theta})
\end{equation}
and second, for the convergence ($\kappa$),
\begin{equation}
	\kappa (\bm{\theta}) = \frac{1}{2} \nabla^2 \psi(\bm{\theta}).
\end{equation}
The convergence is just the surface mass density in the image (or lens) plane at some pointing ($\Sigma$) scaled by the critical surface density of the Universe such that a pointing with 5\% more than the cosmological average yields $\kappa = 1.05$.  In other words,
\begin{equation}\label{eq:kappa}
	\kappa \coloneqq \frac{\Sigma}{\Sigma_\text{crit}},
\end{equation}
with the critical density given by
\begin{equation}
	\Sigma_\text{crit} = \frac{c^2 D_\text{s}}{4 \pi G D_\text{ds} D_\text{d}}.
\end{equation}
It is also related to the deflection potential as $\nabla^2 \psi = 2 \kappa$.

The difference in travel time for this lensed ray and an un-lensed ray is given by the excess time delay,
\begin{equation}\label{eq:excess_td}
    t(\bm{\theta}, \bm{\beta}) =
    \frac{D_{\Delta t}}{c}
    \tau (\bm{\theta}, \bm{\beta}),
\end{equation}
where $c$ is the speed of light and $D_{\Delta t}$ is the time-delay distance (\citealt{refsdal1964}, \citealt{schneider1992}, \citealt{suyu2010}).

The time-delay distance is defined as
\begin{equation}\label{eq:ddt}
    D_{\Delta t} \equiv (1 + z_\text{d})
    \frac{D_\text{d} D_\text{s}} {D_\text{ds}},
\end{equation}
where $z_\text{d}$ is the redshift of the deflector (lens) and $D_\text{d}$, $D_\text{s}$, and $D_\text{ds}$ are the angular diameter distances to the deflector, source, and between the lens and source, respectively. The significance of the time-delay distance is that within each distance term lies an intrinsic $H_0^{-1}$. This leaves 
\begin{equation}\label{eq:H_0_ddt}
    D_{\Delta t} \propto H_0^{-1}.
\end{equation}
We note that $D_{\Delta t}$ has units of distance, and it has a weak dependence on other cosmological parameters.

We assumed the background source and foreground deflector galaxy are sufficiently aligned such that multiple images are observed. As different images require light to travel different paths, their excess time-delays will differ. The time delay between two images, say $\bm{\theta}_i$ and $\bm{\theta}_j$, is thus the difference of their excess time delays:
\begin{equation}\label{eq:tdd}
    \Delta t_{ij} =
    \frac{D_{\Delta t}}{c}
    \left[
        \tau (\bm{\theta}_i, \bm{\beta})
        - \tau (\bm{\theta}_j, \bm{\beta})
    \right] = 
    \frac{D_{\Delta t}}{c} \Delta \tau_{ij}.
\end{equation}

Therefore, constraints on the observed time delay ($\Delta t_{ij}$) and the predicted Fermat potential difference ($\Delta \tau_{ij}$) can be converted into constraints on the time-delay distance ($D_{\Delta t}$), which is an inference on $H_0$ through \Cref{eq:H_0_ddt}.
If a source is variable on sufficiently short timescales, it is possible to measure a time delay, $\Delta t_{ij}$, by extended monitoring of the image fluxes (e.g., \citealt{vanderriest1989, schechter1997, fassnacht1999, fassnacht2002, kochanek2006, courbin2011, bonvin2019}).

The Fermat potential difference, $\Delta \tau_{ij}$, requires three additional parameters for \Cref{eq:tau} (as $\bm{\theta}_i$ and $\bm{\theta}_j$ are simply the positions observed on the sky). Thankfully (neglecting MSD), the source position, $\bm{\beta}$, and the lens potentials at the image positions,  $\psi(\bm{\theta}_i)$ and $\psi(\bm{\theta}_j)$, can be determined by accurately modeling the mass of the system in imaging data; that data we denote by $d_\text{JWST}$. Then, with both $\Delta t_{ij}$ and $\Delta \tau_{ij}$, \Cref{eq:tdd} provides a measurement of $D_{\Delta t}$. With an assumption of a cosmological model, $D_{\Delta t}$ can be turned into an inference on $H_0$.

Thus, the necessary ingredients for cosmographic inference of $H_0$ from strong lensing can be broken into four components. First is the modeling of imaging data, $d_\text{JWST}$, which is the focus of this paper. Second is the measured time delays, $\Delta t_{ij}$, observed from combinations of quasar images ($i$ and $j$). Third is the stellar kinematic information, $\sigma^P$, from spectroscopic data of our lens. The fourth and final ingredient is wide-field spectroscopy and imaging, $d_\text{env}$, which provides environmental information. These are each discussed in part in \Cref{sec:data}.

\subsection{Additional sources of error and degeneracies}
\label{ssec:msd}

\subsubsection{Internal mass sheet degeneracy}

Currently, the primary source of uncertainty in the Fermat potential---and time-delay cosmography at large---is the MSD discussed in \cite{falco1985} (more recently \citealt{wong2020} or \citealt{gilman2020} for substructure discussion or \citealt{birrer2020} for environment discussion). In summary, by adding an infinite ``sheet of mass'' with a constant, uniform surface density, one can find the exact same image positions given a different source position, size, and luminosity, thus completely changing the physical geometry of the system and its associated Fermat potential.

While the time delay ($\Delta t_{ij}$) is an observable, the relative Fermat potential ($\Delta \tau_{ij}$) is not. Instead, the Fermat potential is inferred from constraints on two particular features in the imaging data: the quasar image positions and the extended distortions in the lensed arcs. Consequently, the MSD allows degeneracy in our observed image positions to seep into our Fermat potential, thereby imposing limits on its precision and, via  \Cref{eq:H_0_ddt} , on $H_0$ (e.g., \citealt{falco1985, kochanek2002, saha2006, schneider2013, schneider2014, birrer2016, unruh2017, birrer2021a}).

As the MSD is inherent in the lens model \citep{saha2000, saha2006}, we first note that the MSD is a mathematical degeneracy as opposed to a physical degeneracy. More formally, the MSD is a multiplicative transform of the lens equation that yields the same image positions despite a linear source displacement and transformation of the convergence field, $\kappa$. In fewer words (but more Greek), we first displace the source by some scalar, $\lambda$: 

\begin{equation}\label{eq:beta_transf}
\bm{\beta} \rightarrow \lambda \bm{\beta},
\end{equation}
and we transform the convergence field through
\begin{equation} \label{eq:kappa_transf}
\kappa(\bm{\theta}) \rightarrow \lambda \kappa (\bm{\theta}) + (1 - \lambda).
\end{equation}
Then, solving the lens equation with these quantities, we will find the image positions at
\begin{equation}
\bm{\theta} \rightarrow \bm{\theta}.
\end{equation}
In other words, our direct observable in the imaging data—the image positions—would look the same despite moving the source and changing the convergence field. This is problematic, as the linear source displacement in \Cref{eq:beta_transf} would yield a measurement of
\begin{equation}\label{eq:MSD_H0}
	H_0 \rightarrow \lambda H_0.
\end{equation}
This is currently the leading source of uncertainty in time-delay measurements of $H_0$, and it is broken with the non-lensing constraints on the mass distribution, namely, stellar velocity dispersion \cite{birrer2020,shajib2023}, or by constraints on the absolute luminosity of the source when it is known.
As discussed in the introduction, this paper is concerned with information coming from the lens model itself, and therefore we can neglect the mass sheet transformation while keeping in mind that it needs to be considered when converting the inference from Fermat potential to $H_0$.

\subsubsection{Multiplane lensing}

In addition, the observed time delays will depend on any galaxy (or massive object) having a significant impact on the deflection of our source's light, and these impacts do not need to be at the same redshift as our lens. In this generalized case, we must combine the angular diameter distances of the different lens planes, which is done with the multi-plane lens equation (e.g., \citealt{blandford1986, kovner1987, schneider1992, collett2014, mccully2014}). In our previous assumption, our time delays were proportional to a single unique time-delay distance. Under the multi-plane lens equation, this is no longer true. Thankfully, for quadruply imaged quasars, it very often holds that the mass in a single plane dominates the lensing effect \cite{suyu2020}. This validates the single-lens approximation, as the observed time delays are dominated by the time-delay distance (\Cref{eq:ddt}) at the redshift of the lensing galaxy. This has also been found to be the case for WFI2033-4723 \citep{rusu2020}. This single time-delay distance is referred to as the effective time-delay distance.

\subsubsection{External convergence}
 
An additional correction addressing the lens potential that the light rays experience must be made to validate the single-plane approximation. We must take into account the additional focusing and defocusing that external mass causes, which modifies the measured time delays (e.g., \citealt{seljak1994}). To avoid a biased inference of the time-delay distance, we applied a correction to our modeled value, $D_{\Delta t}^{\text{model}}$, to get the true $D_{\Delta t}$ with
\begin{equation}\label{eq:H0_k_ext}
    D_{\Delta t} = \frac{D_{\Delta t}^{\text{model}}}{1 - \kappa_{\text{ext}}}.
\end{equation}
This is equivalent to adding (or removing) an external sheet of mass in the lens plane to account for the focusing (or defocusing) from the mass unaccounted for in the lens plane. This correction is valid for cases where the line-of-sight (LOS) perturbers are small enough to ignore the effect of higher-order terms \citep{keeton2003, mccully2014}. With \Cref{eq:H_0_ddt}, we note this correction yields a multiplicative factor of $(1 - \kappa_\text{ext})$ for inferences on $H_0$.

However, \Cref{eq:kappa_transf}, informs us that determining the value of $\kappa_\text{ext}$ is generally impossible from the lens model alone due to the MSD. The approach typically taken includes estimating the mass along the LOS combined with the assumption that the deflector's mass profile drops to zero at large radii (e.g., \citealt{fassnacht2006, momcheva2006, momcheva2015, williams2006, wong2010, wells2024a}). Higher-order effects must be explicitly taken into account for perturbers that are extremely massive or particularly close to the lensing galaxy \citep{mccully2017}.

\subsubsection{Mass density slope and time-delay distance degeneracy}

An additional degeneracy exists between the lensing galaxy's radial mass slope and the time-delay distance (e.g., \citealt{kochanek2002, treu2002, wucknitz2002}), even within the context of power-law models. This degeneracy is broken in two ways. First, thanks to JWST, we have imaging of the extended arcs that are barely noticeable in HST data (see \cref{fig:hst_jwst_comp}), whose modeling directly constrains the radial mass slope. Second, we also combine the lensing data with stellar kinematic data (e.g., \citealt{treu2002, koopmans2003, auger2010, suyu2014, yildirim2020}).

\subsection{Analysis of kinematics}
\label{ssec:ana_kinematics}

Stellar kinematics plays a crucial role in mitigating the effects of degeneracies \citep{treu2002, shajib2018, shajib2023, knabel2025a}. With measurements of the stellar velocity dispersion, we can independently probe the 3D mass distribution of our deflector galaxy (after de-projecting from the observed convergence near the center of our lens, $\kappa_\text{obs, cen}$):
\begin{equation}
    \sigma \xrightarrow{\kappa_\text{obs, cen}}  \Phi(r) \rightarrow M,
\end{equation}
where the observed convergence near the center of the deflector galaxy, $\kappa_\text{obs, cen}$, allows us to de-project the velocity dispersion, $\sigma$, to the gravitational potential, $\Phi(r)$.

To this end, we employed the spherical Jeans equation,
\begin{equation}
    \diff{\left( l(r) \sigma_\text{r}(r)^2\right)}{r} + \frac{2 \beta_\text{ani}(r) l(r) \sigma_\text{r}(r)^2}{r} = -l(r) \diff{\Phi(r)}{r},
\end{equation}
with the 3D luminosity density, $l(r)$; the radial velocity dispersion, $\sigma_r(r)$; and the anisotropy parameter that relates $\sigma_\text{r}$ to the tangential velocity dispersion $\sigma_\text{t}$. We defined this anisotropy parameter as
\begin{equation}
    \beta_\text{ani}(r) \coloneqq 1 - \frac{\sigma_\text{t}^2(r)}{\sigma_\text{r}^2(r)}.
\end{equation}

The quantity we actually observed is the luminosity-weighted LOS velocity dispersion. We obtained this from the Jeans equation, as
\begin{equation}
    \sigma_\text{los}^2(R) = \frac{2G}{I(R)} \int^{\infty}_R \mathcal{K}_\beta \left( \frac{r}{R} \right) \frac{l(r) M(r)}{r} \text{d}r,
\end{equation}
with the gravitational constant, $G$; the surface brightness, $I(R)$; and the enclosed 3D mass within a radius $r$, $M(r)$ \citep{mamon2005}. Our function, $\mathcal{K}_\beta$, depends on the parameterization of $\beta_\text{ani}(r)$. We adopted the Osipkov-Merritt parameterization given by
\begin{equation}
    \beta_\text{ani}(r) = \frac{r^2}{r^2 + r_\text{ani}^2},
\end{equation}
with the scaling radius, $r_\text{ani}$ \citep{osipkov1979, merritt1985, merritt1985a}.
This set our function to
\begin{multline}
    \mathcal{K}_\beta \left( u \coloneqq \frac{r}{R} \right) = \\ 
    \frac{u_\text{ani}^2 + 1/2}{\left( u_\text{ani} + 1 \right) ^{3/2}} \left( \frac{u^2 + u^2_\text{ani}}{u} \right) \tan^{-1} \left( \sqrt{\frac{u^2 - 1}{u_\text{ani}^2 + 1}} \right) \\
    - \frac{1/2}{u_\text{ani}^2 + 1} \sqrt{ 1 - \frac{1}{u^2}},
\end{multline}
with $u_\text{ani} \coloneqq r_\text{ani}/R$ \citep{mamon2005}. We could then find the observed aperture-averaged velocity dispersion via
\begin{equation}
    \sigma^2_\text{ap} = \frac{\int_\text{ap} \left[ I(R) \sigma_\text{los}^2(R) \right] \ast \mathcal{S} \ \text{d}x\text{d}y}{\int_\text{ap} I(R) \ast \mathcal{S} \ \text{d}x\text{d}y},
\end{equation}
where we integrated over the aperature, $\int_\text{ap}$, and convolved with the seeing, $\ast \mathcal{S}$.

We note the external convergence modifies our model-predicted velocity dispersion, $\sigma_\text{ap, model}$, through
\begin{equation}
    \sigma_\text{ap, true}^2 = (1-\kappa_\text{ext}) \sigma^2_\text{ap, model}.
\end{equation}

To correct for this effect, we sampled $\kappa_\text{ext}$ and weighted our models by their ability to recover the observed velocity dispersion of the main deflector, $\sigma_\text{obs, D}$ (see \cref{subsec:comb_weight}). 

We note full cosmographic analyses will also feature an additional degree of freedom in models with a $\lambda$ term in this equation. As the purpose of this paper is to evaluate the improvements offered exclusively by imaging data, we focus our efforts on a direct power-law comparison, as imaging alone is unable to directly constrain the MSD.

\subsection{Bayesian analysis}
\label{ssec:ana_bayes}

To predict a time delay under a given cosmology, the lens model infers the Fermat potential difference between two image pairs, which we define as $\Delta \tau$. We aim to describe $\Pr(\Delta \tau \mid O)$, the conditional probability of the Fermat potential difference given our imaging and kinematic observables $O \coloneqq \{O_\text{img}, O_\text{kin}\}$. We do not include the spectroscopic and photometric data used for the lens environment, as the external convergence inference can be accounted for with a separate, explicit prior, $\Pr(\kappa_\text{ext})$.

To provide a full description of $\Pr(\Delta \tau \mid O)$, we must address the functional dependencies of the Fermat potential difference $\Delta \tau (\xi, \kappa_\text{ext})$. First is the external convergence, $\kappa_\text{ext}$, and second is our set of model parameters, $\xi \coloneqq \{\xi_\text{lens}, \xi_\text{light}, r_\text{ani}\}$. Given our observables, we can apply Bayes' theorem to rewrite
\begin{equation}
	\label{eq:bayes_big}
	\begin{split}
		\Pr&(\xi, \kappa_\text{ext} \mid O) \\
		&\propto
		\Pr(O \mid \xi, \kappa_\text{ext}) \times \Pr(\xi, \kappa_\text{ext}) \\
		&=
		\int_2\Pr(O \mid \xi, \kappa_\text{ext}, S, D_{\rm s/ds}) \\
		&\quad \quad
		\times \Pr(\xi, \kappa_\text{ext} \mid S) \dd{S} \dd{D_{\rm s/ds}}.
	\end{split}
\end{equation}
We have introduced $S$, the set of hyper-parameters used by the lens model to describe $O_\text{img}$ (e.g., the number of parameters describing the source, the set of pixels upon which the image likelihood is evaluated), and $D_{\rm s/ds}$, which represents the distance ratio $D_{\rm s/ds} \coloneqq D_{\rm s} / D_{\rm ds}$. We then marginalized on these two parameters.

We note a relatively recent change to this method of Bayesian analysis compared to similar works where the external convergence is now (once again) independent of the model-predicted external shear (see \Cref{subsec:WFDataKext}). This splits the joint likelihood of $\kappa_\text{ext}$ into its own prior. We also separated $O$ into its two independent data components, rewriting \Cref{eq:bayes_big} as
\begin{equation}
	\begin{split}
				\int_2&\Pr(O \mid \xi, \kappa_\text{ext}, S, D_{\rm s/ds}) \\
				&
				\times \Pr(\xi \mid S) \times \Pr(\kappa_\text{ext}) \dd{S} \dd{D_{\rm s/ds}} \\
				&= 
				\int_2\Pr(O_\text{img} \mid \xi, \kappa_\text{ext}, S) \\
				&\quad \times
				Pr(O_\text{kin} \mid \xi,  \kappa_\text{ext}, D_{\rm s/ds}) \\
				&\quad
				\times \Pr(\xi \mid S) \times \Pr(\kappa_\text{ext}) \dd{S} \dd{D_{\rm s/ds}}.
	\end{split}
\end{equation}
Next, we converted the following sub-integral:
\begin{equation}
	\begin{split}
		\int&\Pr(O_\text{img} \mid \xi,  S) \Pr(\xi \mid S) \Pr(S) \dd{S} \\
		&=
		\int\Pr(\xi \mid O_\text{img},  S) \Pr(O_\text{img} \mid S) \Pr(S) \dd{S}.
	\end{split}
\end{equation}
This form is much more convenient, as sampling it gives us the model evidence, $\Pr(O_\text{img} \mid S)$.

\section{Data}
\label{sec:data}

 \begin{figure}[ht]
	 \centering
	 \includegraphics[width=0.5\textwidth]{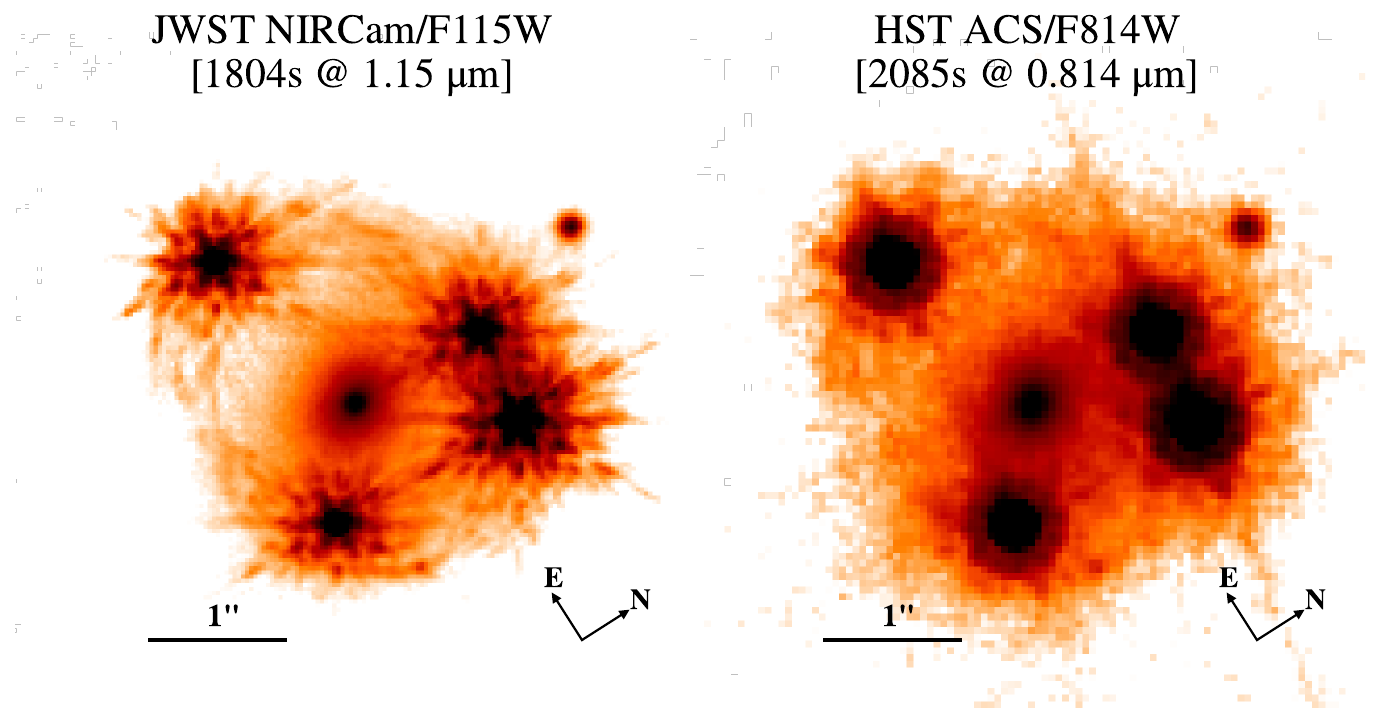}
	 \caption{Comparison of the JWST data used in this analysis, NIRCam F115W, to the HST imaging, ACS/F814W. The ring structure is maximized in each visualization, as it provides the tightest constraints on the deflector's radial mass profile.}
	 \label{fig:hst_jwst_comp}
\end{figure}

As the focus of this paper is to explore the improvements that JWST imaging has to offer to cosmography-grade lens modeling, we used the same data products as the first cosmography model of this system \citep[hereafter H0LiCOW XII]{rusu2020}, replacing the HST imaging with that of JWST.  In this section, we first review the lens system WFI 2033--4723 (J2000: $20^h33^m41.9^s$, $-47^\circ23'43\farcs4$; hereafter WFI2033) in \Cref{subsec:WFI2033}, and this is followed by a discussion of the JWST NIRCam data used in \Cref{subsec:JWSTData}. Next, we summarize the ancillary data, starting with the spectroscopic data used for measurements of velocity dispersions in \Cref{subsec:SpecData}. Finally, we go over the wide-field imaging data and how it is used to constrain the impact of the environment through $\kappa_\text{ext}$ in \Cref{subsec:WFDataKext}.

\subsection{WFI2033--4723}
\label{subsec:WFI2033}

\begin{figure}[ht]
	 \centering
	 \includegraphics[width=0.5\textwidth]{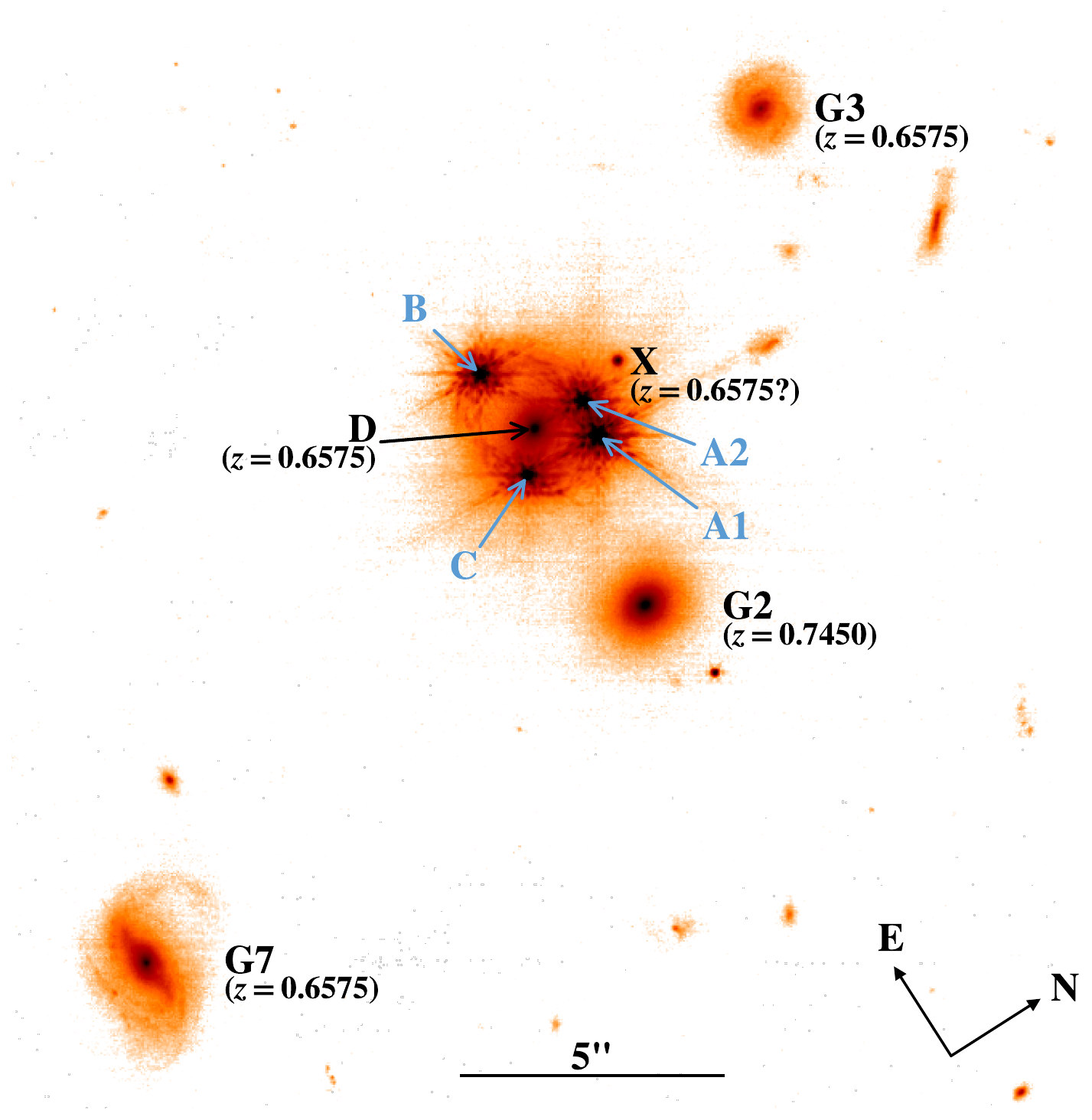}
	 \caption{Environment of WFI2033. Black labels denote the main deflector (D), its nearby satellite (X), and the three galaxy perturbers (G2, G3, and G7) along with their redshifts. The quasar images are labeled in blue.}
	 \label{fig:labeled_environment}
\end{figure}

The discovery of the quadruply imaged quasar WFI2033 was first reported in \cite{morgan2004}, wherein the authors applied color cuts to the Wide-Field Imager (WFI) on the MPG-ESO telescope. The redshift of the source was measured to be $z_{\rm s} = 1.662$ by \cite{sluse2012}, with the deflector at $z_{\rm d} = 0.6575 \pm 0.0002$ \citep{sluse2019} (hereafter H0LiCOW X). Its environment is relatively complicated, with at least six galaxies within $20\arcsec$, three of which need to be explicitly included in the lens model.\footnote{
	The updated analysis in H0LiCOW XII found only one of these three galaxies to be above the standard flexion shift cutoff of $\Delta_3 x > 10^{-4} $ asec, but the two other galaxies were included to be safe. For a fair comparison, we also chose to include all three perturbing galaxies.
} The deflector is also in a group at $z = 0.6588$, with a scale radius predicted to be $r_\text{s,g} = 32.0\pm8.0\arcsec$, which was calculated by assuming its virial mass and radius \citep{rusu2020} (hereafter H0LiCOW XII). In addition, there exists a second group in its foreground ($z = 0.4956$), with its scale radius similarly measured at $r_\text{s,g} = 34.8\pm9.3\arcsec$.

\subsection{JWST imaging}
\label{subsec:JWSTData}

On September 19, 2022, WFI2033 was observed by JWST with both NIRCam imaging and NIRSpec IFU spectroscopy as part of the Guaranteed Time Observing (GTO) program (\#1198; PI: Stiavelli), IFU spectroscopy of the host galaxies of strongly lensed quasars. The program aims to leverage the gravitational lensing effect of the deflector, which spatially stretches the quasar's image, to study the relationship between the quasar and its host galaxy with greater rest frame angular resolution than otherwise possible. Two other quadruply imaged systems were also observed, whose NIRCam models are planned in following publications.

The NIRCam observations were done with four primary \texttt{INTRAMODULEX} dithers with the \texttt{SHALLOW2} readout pattern, producing a slightly higher angular resolution ($\sim\!0\farcs0307$ in the F115W band) than the default pixel size. The bright quasar images were not saturated and as such did not require masking. This is extremely beneficial for our point source modeling, as later discussed in \Cref{sec:psf_modeling}. There were nine groups per iteration, with a total exposure time of $\sim\!1800$ sec.  The uncertainty on the flux in each pixel was estimated from the science image and is a combination of the resampled Poisson, read noise, and flat-field variances added in quadrature. While this data also includes the F150W, F277W, and F356W bands, one of the largest bottlenecks of modeling this high resolution data is computational time. For this reason, we chose to exclusively model the F115W band, as it offers the highest spatial resolution, surpassing HST at the equivalent wavelength. Future analyses will incorporate additional bands as computational efficiency of lens modeling increases, as computing this band alone required three weeks running on 12x36 cores. However, we did test a single model on the F150W band and found consistent results for this system. Our lens model is based on a $4.6''$ (150 pixel) square cutout of the reduced data, with a $0\farcs15$ mask applied to the very center of the deflector.

\subsection{Spectroscopic data}
\label{subsec:SpecData}

Spectroscopic observations were conducted using the ESO-MUSE integral field spectrograph and are presented in H0LiCOW X, which we summarize here. The lens lies within a galaxy group at $z = 0.6588$ with a velocity dispersion of $\sigma = 500 \pm 80  \kms$ measured from 22 member galaxies \citep{wilson2016, momcheva2006, momcheva2015}. The velocity dispersion for the main deflector was measured to be $\sigma_\text{LOS} = 250 \kms$ with a total uncertainty of $\sigma_{\sigma_\text{LOS}} = 19 \kms$. Additional multi-object spectroscopic data were collected for constraints on galaxies in the vicinity ($<2'$) of the lens using the ESO FORS \citep{appenzeller1998} and Gemini GMOS \citep{hook2004} instruments.

JWST NIRSpec data were also taken for WFI2033, which we plan on modeling for spatially resolved kinematics in future analyses, which will enable tight constraints for more general mass models of the deflector. As the focus of this paper is the power-law model, we used the higher resolution of the NIRCam for our modeling..

\subsection{Wide-field photometric data and external convergence}
\label{subsec:WFDataKext}

Measurements utilizing time-delay cosmography are sensitive to whether a system resides in a relatively over- or under-dense LOS (defined as $\kappa_\text{ext}$; see, e.g., \Cref{eq:H0_k_ext}), but these densities are not directly observable. Therefore, $\kappa_\text{ext}$ is estimated by matching observable tracers in the wide-field data to those in simulations—a process that has undergone multiple developments over time \citep{suyu2010b,fassnacht2011,suyu2013,greene2013,wells2023, wells2024a}.
Consequently, wide-field photometric data are needed for photometric redshift measurements, galaxy-star classification, and measurements of the stellar masses of all galaxies $i < 23$ mag within a $120\arcsec$ radius.
These data are described in H0LiCOW X, with the resulting values of $\kappa_\text{ext}$ computed in H0LiCOW XII, which we keep consistent for this analysis.

While the cosmological predictions of WFI2033 require updated sampling of $\kappa_\text{ext}$,\footnote{Recent discussions have raised questions on whether the model-predicted external shear ($\gamma_\text{ext, mod}$) should be included as a prior for solving for $\kappa_\text{ext}$, as it may not correlate one-to-one with the physical shear in these environments \citep[e.g.,][]{etherington2023}. As this topic is not the focus of this work, we decided to sample the same distribution of $\kappa_\text{ext}$—still informed by $\gamma_\text{ext, mod}$—as H0LiCOW XII to guarantee a fair comparison.} this has a negligible impact on a one-to-one comparison of lens models due to the resulting model independence of $\kappa_\text{ext}$. Because of this, any comparison of models must be done using the same sample of $\kappa_\text{ext}$. In this work, our aim is to compare models of HST- and JWST-based data. So, for a fair comparison, we used the same sample of $\kappa_\text{ext}$ as \citet{rusu2020} in our lens modeling, and we note that any inference on cosmological parameters will require an update in the sampling of external convergence, which is left for future work.

\section{JWST PSF reconstruction}
\label{sec:psf_modeling}

Careful consideration must be taken to ensure the complex, extended PSF structure does not bias measurements of our parameters. \citet{shajib2022} and \citet{ding2021} demonstrated that, even for the same imaging data, models with different PSFs can have significant discrepancies on crucial model parameters, such as the power-law slope, $\gamma$, or the external shear, $\gamma_\text{ext}$. As this is the first cosmographic strong lens model with JWST's challenging PSF, we discuss a new method to model the PSF and compare it to the method adopted by recent studies to describe the HST PSF.

While the extent of JWST's PSF may be similar to that of HST, Figure~\cref{fig:hst_jwst_comp} illustrates how the complexity of the PSF has increased. A crucial step in lens modeling is the separation of the source's light into the extended component, which contributes to the ring-like structure of the lens, and the point source component, which contributes to the bright images seen at image positions $\text{A, B, C, and D}$. A model incorrectly disentangling these components can fit a nonphysical--or even non-existent--light structure, potentially biasing the model. For example, a bias in the power-law slope closely tied to the source light profile will bias the final inference of $H_0$.

\subsection{Initial PSF model}
\label{subsec:mod_init_psf}

To generate an initial PSF, we followed the standard procedure of extracting stars in the field as reference. Ideally, a selection of stars should be as close to the lens galaxy in angular separation as possible to minimize the effects of PSF distortion across the field. However, a larger constraint on star selection proved to be finding stars of similar flux to the lensed quasar images. Having similar flux is crucial to avoid discrepancies caused by saturation and nonlinear effects at the bright end and by noise at the faint end. Across the entire $2.5'$ square field, 13 stars have desirable flux, with three saturating the centers of their PSFs and four being on the edge of an exposure. Our selection is composed of the six remaining stars.

Previous strong lens modeling approaches have typically utilized either the \texttt{astropy}\footnote{http://www.astropy.org} core package \texttt{photoutils} \citep{collaboration2022} or the more recently developed \texttt{PSFr} \citep{birrer2022}. We compare our PSF modeling technique to \e{PSFr}, which implements the following algorithm:

\begin{enumerate}
	\item Obtain an initial guess of the PSF by stacking the cutout stars (without any recentering).
	\item Calculate the sub-pixel centroid of each star's cutout.
	\item Fit for the sub-pixel centroid of the PSF model.
	\item Line up the sub-pixel centroids of your PSF model and each of the cutout stars.
	\item Interpolate the PSF model ``into the frame" of each star with sub-pixel interpolation (keeping their centroids lined up).
	\item Calculate the residuals of each interpolated PSF model to its star's cutout data.
	\item To combine the residuals from each of these different sub-pixel-interpolated frames, transform these residuals back ``into the frame" of your PSF model (an ``inverse sub-pixel shift").
	\item With all the residuals in the PSF model's frame, stack them (e.g., median) to inform a correction to your PSF model.
	\item Repeat steps (4)-(8), optionally repeating step (3).
\end{enumerate}

In this work, we implement \e{STARRED} \citep{michalewicz2023, millon2024} as an alternative method to generate an initial PSF for strong lens modeling. \e{STARRED} directly addresses a few of the novel challenges that JWST proposes for strong lens modeling in particular.

The first benefit of \e{STARRED} is wavelet regularization, which increases the population of suitable stars for PSF generation by reducing the influence of varying noise levels among the stars on the resulting PSF model. Instead of representing our PSF in a spatial pixel basis, \e{STARRED} exploits the isotropic wavelet basis of starlets \citep{starck2006}. Our PSF is sparsely represented in this starlet space, meaning it can be reconstructed with a linear combination of very few of these starlet bases. This is not true of the noise. As a result, one can simply select and add up the few coefficients required to reconstruct the PSF, dropping the low-power coefficients composing noise. This denoising process is called image thresholding, and it is one of the main benefits of wavelet regularization.
Second, sparse regularization has been shown to improve photometric accuracy and astrometric precision to better than a hundredth of a pixel for high signal-to-noise data \citep[Fig. 4 of ][]{millon2024}, the latter of which is crucial for tight constraints on image positions and thus $H_0$  \citep{birrer2019}.

We estimated the uncertainty in the PSF based on residuals of our initial fitting of field stars. Residuals without the additional uncertainty due to the PSF are shown in \Cref{fig:psf_error_residuals} and discussed in \Cref{ssec:psf_noise_estimation}. We additionally cropped the initial PSF models based on a minimum threshold, removing the bottom $\sim\!1\%$ of the initial kernel. This removed the extended background noise from the initial PSFs (due to the differing fluxes of the quasar versus the stars used in the initial fit). Without this additional cropping, the point-sources used in the modeling would have extended light profiles that add a nonphysical sheet of flux across the image.

\subsection{Iterative PSF updating}
\label{subsec:mod_iter_psf}

\begin{figure}[ht]
	\centering
	\includegraphics[width=0.5\textwidth]{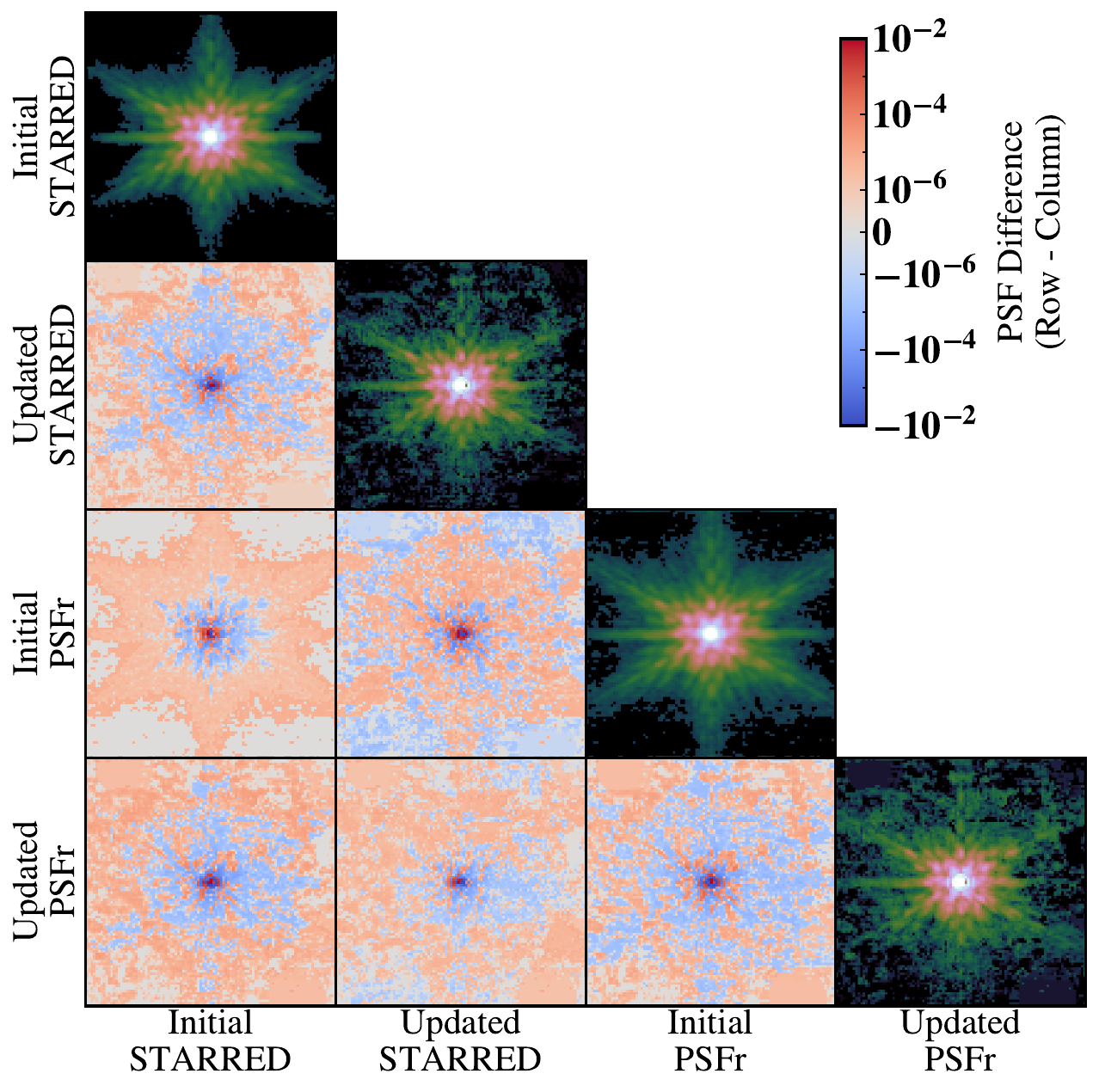}
	\caption{Comparison of the PSF kernels used. First we show the STARRED initial PSF kernel, and it is followed by an example of an updated PSF from one of the STARRED models. Then we show the PSFr initial PSF kernel, which is again followed by an example of an updated PSF from a PSFr model.}
	\label{fig:psf_kernel_comparison}
\end{figure}

After acquiring our initial PSF models, we iteratively updated each model based on the quasar images during the image forward modeling (result shown in \cref{fig:psf_kernel_comparison}). By first subtracting the best lens light and extended source surface brightness profiles, we could update the PSF with the additional four remaining quasar point sources, which was repeated until the PSF model converged to a solution (or reached 100 iterations). This entire process was repeated six times during the initial fitting process for each model in order to avoid biasing the PSF model based on our initial parameter values or overall model parameterization. We used the same PSFr-based iteration method for the two initial PSF models. We used a completely asymmetric PSF, as we found the image residuals cannot be accurately modeled with any amount of symmetry. We did not update our PSF error map during the fitting (for more information on the error map, see \Cref{ssec:psf_noise_estimation}).

\section{Modeling}
\label{sec:mod}

 \begin{figure*}[ht]
	\centering
	\includegraphics[width=\textwidth]{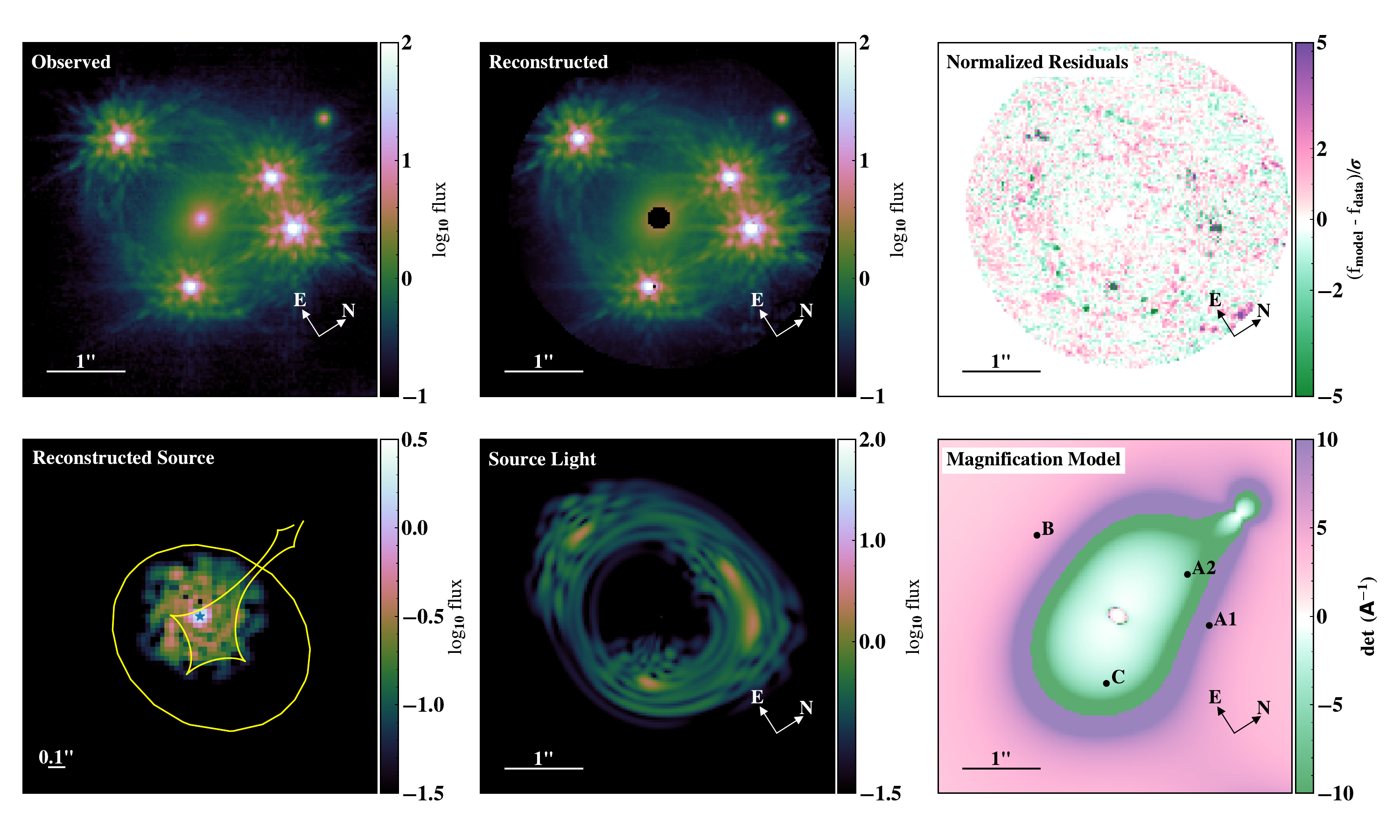}
	\caption{Example model output from the best performing model shown in \Cref{tab:res_mods}. We note that the other models are qualitatively indistinguishable. The top-left panel shows the observed image, the top-middle panel shows the model-predicted reconstruction of the image, and top-right panel shows the fit's residuals normalized by the estimated uncertainty of each pixel (see \cref{ssec:psf_noise_estimation} for more details). At the bottom left is the reconstructed source plot, featuring more detail than any parametric source modeled for time-delay cosmography thus far. The star symbol denotes the location of the quasar host galaxy's centroid. The bottom-middle panel shows the lensed unconvolved extended-source light, and the bottom-right panel shows the magnification map of the system.}
	\label{fig:example_fit}
\end{figure*}

In this section, we describe the modeling procedures used to fit the JWST NIRCam data for the final inference on lens parameters. For our analysis, we used \texttt{Lenstronomy},\footnote{https://github.com/sibirrer/lenstronomy} a python software package developed by \cite{birrer2018} \citep{birrer2015, birrer2021}. The accuracy of \texttt{Lenstronomy} has been verified through methods such as the Time-Delay Lens Modelling Challenge, where ``true parameters" of a simulated system were hidden from different analysis teams \citep{ding2021}.  Since then, it continues to be used as one of the primary modeling methods for time-delay cosmography today.
The priors on all parameters are uniform, and their values are based on initial modeling of the system, unless otherwise stated, and are summarized in \Cref{tab:components_parameters}.

\subsection{Main deflector galaxy D}
\label{subsec:mod_D}

The main deflector is a massive elliptical galaxy. Following standard practice and to facilitate a comparison with the previous analysis \citep{rusu2020}, we characterized its mass profile via the power-law elliptical mass distribution (PEMD; \citealt{barkana1998}) specified by
\begin{equation}
	\kappa_\text{PEMD}\left(\theta_1 , \theta_2\right) \coloneqq \frac{3 - \gamma}{2}\left[\frac{\theta_\text{E}}{\sqrt{q_\text{m}\theta_1^2 + \theta_2^2/q_\text{m}}}\right]^{\gamma-1}.
\end{equation}
The free parameters are the logarithmic slope ($\gamma$), the Einstein radius ($\theta_\text{E}$), and the axis ratio ($q_\text{m}$). If the angle of the major axis is given by $\varphi_\text{m}$ with reference to the RA-Dec frame, then we define the coordinates, $\left(\theta_1, \theta_2\right)$, aligned with the major and minor axes. 

For the deflector's light profile, we used two (elliptical) Sérsic profiles,
\begin{equation}
	I(\theta_1,\theta_2) = A \exp{
        -k \left( \left( \frac{\sqrt{q_L \theta_1^2 + \theta_2^2 / q_L}}{r_{\text{eff}}}\right)^{1/n} - 1 \right)
		 },
\end{equation}
with the amplitude ($A$), the effective (half-light) radius ($r_\text{eff}$), the axis ratio ($q_\text{L})$, the Sérsic index ($n$), and a corrective constant ($k$) \citep{sersic1968}. 

The two Sérsic light profiles struggle to fit the very center of the deflector, but this has little impact on our lens model. Since fitting the lens light serves two roles—providing a light model for the kinematic analysis and removing the lens light contribution from the extended source's ring structure—we assessed how a mask would impact these goals. Preliminary tests showed that masking this region has little effect on model-predicted velocity dispersions, and the ring structure lies sufficiently far from the deflector's center. Therefore, we masked the central $0\farcs15$ to prevent this region from driving the fit (shown in \Cref{fig:example_fit}).

We included an additional external shear term to account for distortion by the large-scale structure of the lens environment, such as the group's halo. We further constrained the ellipticity and position angle of the mass profile based on the observed ellipticity and position angle of the light profile following \citet{schmidt2022}. To mitigate degeneracies caused by the external shear, we also applied a Gaussian prior on the ellipticity parameters.

\subsection{Quasar and host galaxy}
\label{subsec:mod_quasar}

The quasar was modeled as point sources in the image plane convolved with the PSF, whose generation is further detailed in \cref{sec:psf_modeling}. We constrained the image positions within an astrometric precision of $0.34 \, \text{mas}$, which is well below the astrometric requirements for cosmography \citep{birrer2019}.

In addition to the point-source structure generated by the quasar, we also fit the extended light produced by the quasar's host galaxy. To account for its complexity, we added standard linear shapelet functions \citep{refregier2003, birrer2015} to our quasar host galaxy's Sérsic light profile. The host galaxy's parameterization was explicitly left scale invariant, as forcing a fixed source reconstruction scale can cause an incorrect solution to the mass-sheet transformation to be assumed and thus uncertainties in $H_0$ would be underestimated \citep{birrer2016}.

We found the optimal number of shapelet bases to reconstruct the source by finding the Bayes information criterion (BIC) turnover point, where the additional complexity of more shapelets becomes incommensurate with the goodness-of-fit as determined by the BIC. All initial conditions were kept the same between these models, with the only distinction being the maximum polynomial order permitted by the model. Notably, we find that the largest polynomial order tested ($n_\text{max}=30$) still recovers better BIC values than lower-order models; however, the computational time is nonlinear with $n_\text{max}$. Because of this, we adopted the significantly faster polynomial orders of $n_\text{max}\in\{18,20,22\}$ despite a minor drop in the accuracy of the source reconstruction.

\subsection{Galaxy-scale perturbers}
\label{subsec:mod_pert}

Of the nearby galaxy perturbers, the authors of H0LiCOW X determined three had a flexion shift large enough such that their direct inclusion in the lens model is necessary due to potential higher-order effects. The authors of H0LiCOW XII redid the flexion shift analysis and found only one of these perturbers (labeled G2) to be above the threshold; however, they included the other two perturbers (labeled G3 and G7) to be safe.\footnote{In addition to galaxies, the authors of H0LiCOW X also measured and made predictions on the impact of galaxy groups on the lens potential. While no group was found to have a flexion shift above the typical threshold of $\Delta_3 x > 10^{-4}$, the authors of H0LiCOW XII also included two group halos in the model for extra caution. While these models did recover the best BIC values, the improvement was within the typical BIC variance. Because of this, we do not explicitly include these groups in the model in order to minimize the number of nonlinear parameters and the computational time.}. In addition, there is a small perturber near the main deflector, which is assumed to be a satellite at the same redshift due to a lack of a spectrum (labeled X in \cref{fig:labeled_environment}).

The perturber G2 has the largest flexion shift of the galaxy perturbers due to its proximity and comparable size to the main deflector D. We also included a light profile for G2 due to its proximity, representing it with a single Sérsic. G2 also lies outside of the lens plane, at a redshift of $z_\text{G2}=0.7450$ (while $z_\text{D}=0.6575$). We took this into account when converting its velocity dispersion ($\sigma_\text{LOS, G2} = 223 \lowup{8}{14} \kms$) into the Einstein radius, assuming that the lensing velocity dispersion is equal to the stellar velocity dispersion within the scatter \citep{treu2006}. Thus, we effectively modeled the system with the multi-plane lens equation.

The perturber G3 lies significantly far ($\approx 7.2 ''$) from the lens as to not require modeling of its surface brightness profile. However, G3 may have a flexion shift large enough to require direct modeling of the mass distribution. Therefore, we modeled its mass with an SIS profile. In H0LiCOW X, a velocity dispersion of $\sigma_\text{LOS, G3} = 79 \lowup{19}{23} \kms $ was measured, which we converted to a prior on the Einstein radius with the same redshift as the deflector, as their redshifts are consistent within the range allowed by peculiar velocities.

Similarly, the perturber G7 is also far enough away such that a surface brightness profile does not need to be included. However, its gravitational effects may not be negligible, and therefore we modeled G7 with an SIS profile anchored to its measured velocity dispersion of $\sigma_\text{LOS, G7} = 166 \lowup{6}{7} \kms$ at the redshift of the deflector.

\subsection{Satellite X}
\label{ssec:mod_X}

In the previous HST power-law models, the authors of H0LiCOW XII found the mass of the satellite X to be unconstrained by the imaging data, with an SIS Einstein radius of $\theta_\text{E} = 0.001\lowup{0.001}{0.001}$. In this work, we use the same mass and light profiles with SIS centered on a Sérsic, due to its proximity to the deflector. To get a rough initial estimate of the satellite's Einstein radius, we assumed it follows the Faber-Jackson relation \citep{faber1976} and estimated its radius based on a preliminary model's radius of the deflector, yielding $\theta_\text{E,X, init} \approx 0.15 ''$.

\subsection{Flexion}
\label{subsec:mod_flexion}

Models for time-delay cosmography feature higher-order corrections of the lens potential for any mass affecting the system that is not explicitly included in our lens models. The second-order corrections for an external mass' effect on the lensing potential are given by the external convergence correction (\Cref{subsec:WFDataKext}), which changes the scale (or magnification) of the lensed image, and the external shear term (\Cref{ssec:external_shear}), which stretches the lensed image.

One of the questions for higher spatial resolution data is whether these typical corrective terms used in the past are sufficient for this data or if the increase in resolution requires additional higher-order corrections. After the second-order shear and convergence terms, the next term would be the third-order correction called flexion \citep{goldberg2005}. To test this additional correction, one of our systematic choices is the addition of a free flexion term, which allowed the model to additionally vary a triangular bending effect on the lensed light. This additional term has only been applied to one time-delay cosmography system thus far (J1721+8842, Schmidt et al., in-prep).

\subsection{Additional constraints}
\label{subsec:additional_constraints}

The velocity dispersions of the galaxy-scale perturbers (G2, G3, and G7) are used to estimate their Einstein radii. The ratios of these radii are fixed and scaled together using a single scaling parameter. To encode the maximum amount of information, we converted the lower and upper $1\sigma$ estimates of the velocity dispersions into our $1\sigma$ estimates in $\theta_\text{E}$. This mitigates potential degeneracies arising from the perturbers’ positions (and the additional external shear) and prevents the model from optimizing the radii in a way that is inconsistent with their measured redshifts and velocity dispersions.

To accurately account for G2, we had to solve the multi-plane lens equation, which requires adopting a cosmology.  Since we are concerned with comparing two lens models and not with doing cosmographic inference, this does not affect our conclusions.

We joined the source-plane position of the quasar with the center of the Sérsic and its shapelets, which comprise the host galaxy's extended light. We also joined the centers of the two Sérsics describing the bulge and disk components of D. For both the satellite and G2, we bound the center of their mass profiles with their light profiles.

The point-source positions of the quasar were constrained by the pixels of the JWST image, assuming the PSF model described in \cref{sec:psf_modeling}. We efficiently solved the lens equation to constrain six of the lens mass parameters, one for each of the independent, relative positions of the four images. This was done with the \texttt{PROFILE\_SHEAR} solver in \texttt{Lenstronomy}, which solves for the following nonlinear parameters of D: its centroid (RA and Dec), axis ratio and position angle (represented as $e_1$ and $e_2$), Einstein radius ($\theta_\text{E}$), and the angle of external shear ($\phi_\text{ext}$).

\subsection{Image forward modeling}
\label{subsec:fm}

Given a proposed set of these parameters, the linear response functions in the data (such as amplitude parameters, point sources, and coefficients of shapelets) are rendered in the image plane and optimized with a linear minimization method based on the imaging likelihood.
We supersampled the central regions of the point sources by a factor of three for ray-tracing and performed the PSF convolution on this subset of the data using the \texttt{adaptive} compute mode. This provides improved constraints on the image positions and avoids aliasing and other artifacts resulting from modeling sub-pixel features in the PSF. We did not apply the PSF convolution on the entire supersampled image, as this quickly dominates the computational cost. To benefit from the supersampled grid, the PSF kernel must also be supersampled at a scale of three. Without a supersampled PSF, the sub-pixel scale features of the PSF would not be accurately accounted for in the convolution, leading to poor fitting of the image positions. By confining this process to only the regions nearest the PSFs (radius of $\sim 0.35''$), we still gained the increase in accuracy without suffering the computational cost.

Our uncertainty in the data is given by the 2D resampled Poisson, the readout noise, and the flat-field variance estimates summed in quadrature, which we then scaled by the weight map to account for the drizzling procedure. We added additional uncertainty in the data to account for uncertainty in the PSF models (discussed in \cref{ssec:psf_noise_estimation}).

\subsection{Likelihood sampling}
\label{subsec:comb_like}

When computing the image likelihood, we penalized models that failed to map the images back to within $0\farcs004$ of each other in the source plane. We also punished the models with negative Sérsic or point-source amplitudes. We only computed the image likelihood on a subset of the data, as light at the center of the main deflector or far outside the ring contains little information as long as the light profile of the main deflector is accurately represented near the ring itself. The \texttt{MASK} systematic test (in \cref{subsec:comb_sys}) extends the size of the outer radial mask to ensure this choice does not bias our results.

We introduced a likelihood penalty for models where the main deflector's mass profile is significantly more elliptical than or misaligned with its light profile. This was inspired by the approach of \citealt{schmidt2022}. Without this additional constraint, we find the large number of perturbers lead to a nonphysical degeneracy in the deflector's mass ellipticity. We added a flat prior requiring
\begin{equation}
    q_\text{LL} - q_\text{L} < 0.2,
\end{equation}
i.e. that the mass profile axis ratio $q_\text{L}$ agrees within 0.2 of the lens-light's axis ratio $q_\text{LL}$.
Similarly, we enforced a constraint to align the position angles of the mass profile, $\phi_\text{L}$, and the lens-light profile, $\phi_\text{LL}$. Specifically, we penalized configurations where the angular difference between these two position angles exceeds a certain threshold, which is a function of the mass profile's axis ratio, $q_\text{L}$. The alignment criterion was defined as
\begin{equation}
\Delta \text{PA} = |\phi_\text{L} - \phi_\text{LL}| \times \frac{180}{\pi},
\end{equation}
where $\Delta \text{PA}$ is the difference in position angles in degrees. Configurations were rejected if
\begin{equation}
\Delta \text{PA} > 10 - \frac{5}{q_\text{L} - 1.0001}.
\end{equation}
This condition ensures that the mass and light profiles remain well-aligned, particularly for more elliptical systems, where misalignment is less tolerable. We introduced a small offset to prevent numerical instabilities in the computation.

\section{Combined analysis}
\label{sec:comb}

In this section, we describe how the systematic tests were combined to form the final inference on the Fermat potential. We first reviewed the systematic tests considered for the final inference in \cref{subsec:comb_sys} (a complete list of preliminary tests can be found in \cref{subsec:appendix_additional_sys_tests}). Then, we discuss how these models were weighted and combined in \cref{subsec:comb_weight}.

\subsection{Primary systematic tests}
\label{subsec:comb_sys}

Here, we summarize the model combinations introduced in \cref{sec:mod}:

\begin{itemize}
    \item Two choices for the inclusion of free flexion:
    \begin{itemize}
        \item \e{No Flexion}
        \item \e{Flexion}
    \end{itemize}
    \item Three choices with increasing flexibility in source surface brightness complexity\footnote{We verified our choices are above the minimum data supported threshold in \cref{ssec:source_recons_supersampling}.}:
    \begin{itemize}
        \item \e{SERSIC+18n$_\text{max}$}
        \item \e{SERSIC+20n$_\text{max}$}
        \item \e{SERSIC+22n$_\text{max}$}
    \end{itemize}
    \item Two different pixel masks for the imaging likelihood:
    \begin{itemize}
        \item \e{MASK 2.1''} (or 68 pixel) radius
        \item \e{MASK 2.2''} (or 71 pixel) radius
    \end{itemize}
    \item Two different initial PSF generation methods:
    \begin{itemize}
        \item \e{STARRED PSF}
        \item \e{PSFr PSF}
    \end{itemize}
\end{itemize}

This gives 24 separate model configurations—and thus 24 separate Fermat potential difference posteriors—which are combined through a weighted sampling (discussed in \cref{subsec:comb_weight}). To optimize the lens model, we first used a particle swarm optimization (PSO) algorithm to identify high-likelihood regions, then refine those solutions with a Markov chain Monte Carlo (MCMC) sampler to fully map out the posterior distributions. We also iteratively refined the PSF for each model between the PSO steps, making it more robust to any bias introduced by model choices. Each model takes approximately one week to run on 36 cores.

\subsection{Model weighting}
\label{subsec:comb_weight}

To ensure our parameterization does not bias our results, we gave each model a weight based on its ability to predict the data and then combined these weighted models to form the final inference. As expected, we observed significant variability in the ability of the models to reconstruct the imaging data. Further, goodness-of-fit statistics on imaging data alone cannot distinguish models with diverging physical interpretations, as profiles can become mass-sheet transforms of each other (\cref{ssec:msd}). To address this, we assigned the models two separate weights that are combined for the final inference: The first reflects their ability to reconstruct the imaging data, and the second is related to their ability to predict the deflector's velocity dispersion. The models were then combined to form a final weighted prediction. The resulting model weights can be found in \Cref{tab:res_mods}.

\subsubsection{Imaging-based weighting}
\label{ssec:imaging_based_weighting}

To prevent overly complex models from being unjustly favored, we employed the BIC to statistically weight our 24 models. Compared to similar criteria, such as the Akaike information criterion, BIC is more penalizing to model complexity, thus encouraging more parsimonious solutions. Formally,
\begin{equation}
    \mathrm{BIC} = \ln(n)\,k \;-\; 2\,\ln(\hat{L}),
\end{equation}
where $n$ is the number of data points, $k$ is the number of model parameters, and $\hat{L}$ is the maximum likelihood of the model. Within a $2.1''$ mask,  $n = 14,592$ pixels; and for $2.3''$, $n = 15,930$ pixels. The number of nonlinear parameters is either 32 or 36, based on choice of Flexion, while the number of linear parameters ranges from 199 to 285. The BIC remains a valid criterion here as $n \gg k$.

Formally, the BIC comparison is only valid for models fit to the same data set. One of our systematics is mask size, which inherently changes the size of the data. As an approximation, following \citet{wong2017} and \citet{birrer2019a}, we used the same mask (the smaller of the two) to calculate $\chi^2$ and the BIC, allowing consistent model comparisons despite slight differences in the masked region.

When comparing any two models, $M_1$ and $M_2$, with BIC values, $\mathrm{BIC}_1$ and $\mathrm{BIC}_2$, their relative probabilities can be written as
\begin{equation}
    \frac{p(M_1)}{p(M_2)} \;\propto\; \exp\!\left[-\frac{\mathrm{BIC}_1 - \mathrm{BIC}_2}{2}\right].
\end{equation}
Defining $\mathrm{BIC}_\mathrm{min}$ as the lowest BIC among all models, we weighted each model, $n$, by
\begin{equation} \label{bic_weight}
    f_{\mathrm{BIC}}(x) =
    \begin{cases}
        1 & \text{if } x \le \mathrm{BIC}_\mathrm{min},\\[6pt]
        \exp\left[-\frac{x - \mathrm{BIC}_\mathrm{min}}{2}\right] & \text{if } x > \mathrm{BIC}_\mathrm{min},
    \end{cases}
\end{equation}
where $x$ is the model’s BIC. This piecewise definition penalizes large BIC values.

However, following \citet{birrer2019a} and TDCOSMO~IX, we additionally convolved each model’s nominal $\mathrm{BIC}_n$ with a Gaussian of width $\sigma_\mathrm{intrinsic}$, reflecting our uncertainty in BIC due to limited sampling and systematics. We estimated $\sigma_\mathrm{intrinsic}$ by comparing BIC values across our shapelet tests ($n_\text{max}={18, 20, 22}$) while fixing the other tests (\e{No Flexion PSF}, \e{STARRED PSF}, \e{MASK 2.1''}). That convolution ensured that small fluctuations in $\mathrm{BIC}_n$ do not spuriously dominate the weighting. Specifically, each model ($n$) received a single BIC-based weight:
\begin{equation}
    W_{\mathrm{abs},n} = \int_{-\infty}^{\infty} \frac{1}{\sqrt{2\pi}\,\sigma_\mathrm{intrinsic}}
    \exp\left[-\frac{\bigl(\mathrm{BIC}_n - x\bigr)^{2}}{2\,\sigma_\mathrm{intrinsic}^2}\right]
    f_{\mathrm{BIC}}(x)\,dx.
\end{equation}
The final relative weight, $W_n$, was then normalized so that $\max(W_n) = 1$.

Because each model has exactly one BIC-based weight, all individual MCMCs drawn from that model share that weight. Models with a lower (better) BIC value received proportionally more draws in the final aggregated parameter space. We find that resampling the MCMC chain for each model according to $W_n$ offers an efficient combination of our systematically different lens models, ensuring each model is represented fairly in the final posterior distribution.

As our models are a relatively sparse sample of all possible configurations, the BIC-based weights help mitigate overfitting tendencies while still including sufficiently flexible models. For parameters such as the Fermat potential differences, our sampling is dense enough that the weighted posterior remains robust. For the broader BIC distribution itself, the convolution integral accounts for scatter, ensuring no single model with an artificially low BIC is overemphasized.

\subsubsection{Kinematics-based weighting}
\label{ssec:kinematics_based_weighting}

Our second weight assesses consistency with the observed velocity dispersion of the main deflector. For each MCMC sample, we computed the model-predicted velocity dispersion, $\sigma_\text{ap, model}$, and then assigned a kinematic weight based on a Gaussian likelihood centered on the observed velocity dispersion, $\sigma_\text{obs,D}$:
\begin{equation}
f_\text{kin} = \mathcal{N}\!\bigl(\sigma_\text{ap, model} \mid \sigma_\text{obs, D},\,\sigma_\text{obs,$\sigma$}\bigr),
\end{equation}
taking into account the uncertainty in the observed velocity dispersion, $\sigma_\text{obs,$\sigma$}$. This effectively penalizes models that predict a significantly different $\sigma_\text{ap, model}$ from the measured dispersion.

We then combined the kinematic weight with the BIC-based imaging weight to form a ``total weight":
\begin{equation}
W_{\mathrm{total}} \;=\; W_{\mathrm{BIC}} \;\times\;f_\text{kin} .
\end{equation}
After computing $W_{\mathrm{total}}$ for each model sample, we normalized it such that $\max(W_{\mathrm{total}}) = 1$. For the final posterior combination in the corner plots, each MCMC sample was drawn in proportion to $W_{\mathrm{total}}$. For the final parameter estimates, we instead calculated the weighted median and the 16th and 84th percentiles.

\section{Results and discussion}
\label{sec:results_and_discussion}

\begin{figure*}[ht]
	\centering
	\includegraphics[width=\textwidth]{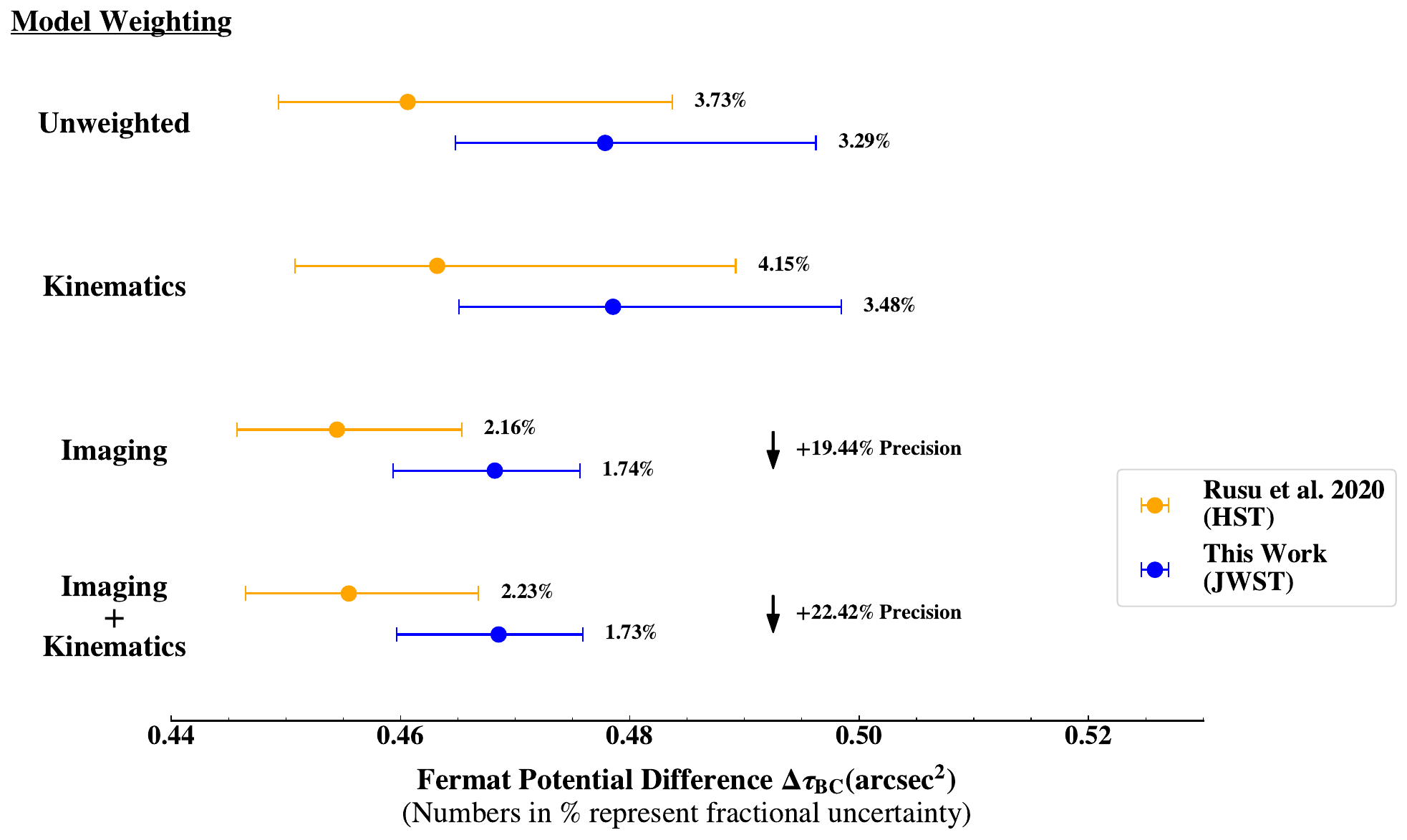}
	\caption{Comparison of the Fermat potential differences of images B and C between the HST models \citep[][orange]{rusu2020} and this work (blue). The fractional uncertainty (measured as the standard deviation over the median) is given to the right of the measurements. Vertically, we display the different weighting schemes, starting with the unweighted models at the top. In the second row, we have the kinematic weighting, which preferentially weights both estimates slightly higher and wider. This is due to the model-predicted velocity dispersions being lower than the spectroscopically observed velocity dispersion measured in H0LiCOW X. However, the velocity dispersion estimates are consistent across the HST and JWST models (see \Cref{fig:corner_plot_final}). The imaging weighting in the third row is far more constraining than the kinematics for this system, and it indicates that the best-fitting models tend to be on the lower end of the models tested. This holds for both HST and JWST results. The estimate recovered by our best-fit models is $\sim\!3.1\%$ higher, most likely due to tighter constraints on the satellite's mass provided by the NIRCam imaging. The last row combines both kinematics and imaging weights for the final estimate, with a $22\%$ increase in precision compared to HST imaging.}
	\label{fig:FPs_comparison_HST_JWST}
	\end{figure*}

We recovered a $22\%$ tighter estimate of the Fermat potential difference after incorporating both kinematic and imaging constraints (\Cref{fig:FPs_comparison_HST_JWST}). While our results remain consistent with previous models, our measured $3.1\%$ increase in the Fermat potential suggests a corresponding $3.1\%$ increase in the inferred value of $H_0$ by this lens, assuming all other parameters remain unchanged. This adjustment would shift the median measured $H_0$ for this lens from 71.6 to 73.8 kms$^{-1}$ Mpc$^{-1}$, with uncertainties likely similar to the HST result. We find a similar trend in the Fermat potential difference between images B-A1 and B-A2. A comprehensive cosmographic analysis, including a resampling of the external convergence ($\kappa_\text{ext}$), will be presented in a forthcoming paper, which could further adjust the overall $H_0$ estimate in either direction.

\begin{figure*}[ht]
	\centering
	\includegraphics[width=\textwidth]{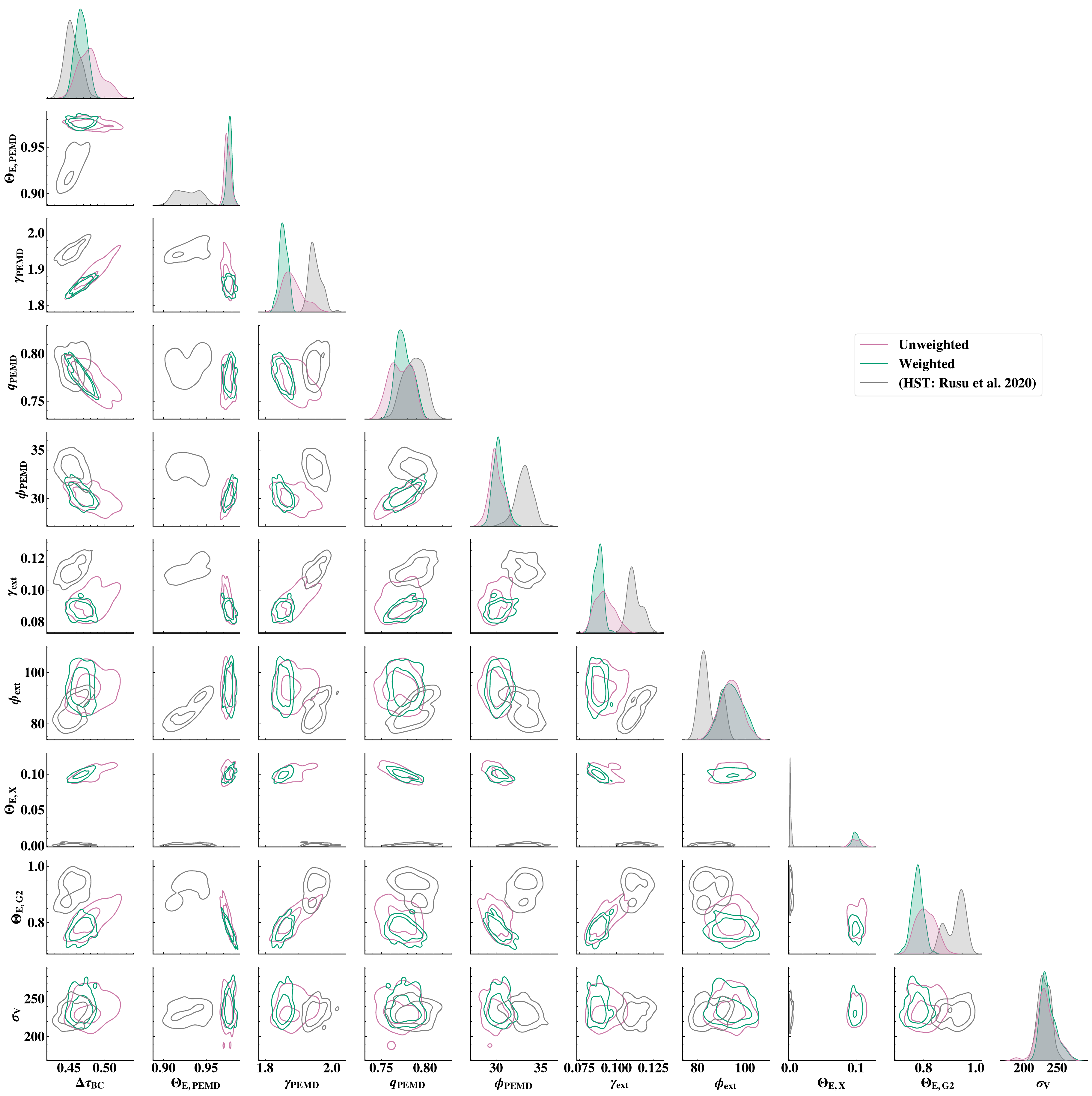}
	\caption{Parameter distributions from our lens model results compared to H0LiCOW XII (HST-based data) in gray. Unweighted samples are shown in pink, with the final BIC+Kinematic-weighted selection shown in green. The BIC-only weighted sample is nearly indistinguishable from the BIC+Kinematic weighted sample, implying the difference in predicted kinematics between models is negligible, and so it is not included in the figure. The contours represent the 68.3\%, and 99.7\% quantiles.}
	\label{fig:corner_plot_final}
	\end{figure*}

While the Fermat potential differences between the images are consistent with those of the HST models, we noted significant differences in the Einstein radius and the power-law slope. The primary source of this discrepancy is the handling of the perturbers, specifically the satellite X and nearby galaxy G2. Due to their placement relative to the deflector, a degeneracy occurs that allows the model to mimic the effect of satellite X's mass perturbation. By increasing the ellipticity of galaxy G2, increasing the shear, and adjusting the Einstein radius and slope of the main deflector, the HST models were able to mimic the convergence at the image positions to get nearly the same result as when the satellite is properly constrained. This is discussed further in \Cref{sec:XG2}.

We find models including free flexion introduce a new degeneracy between the Fermat potential and the Einstein radius of the main deflector, which the non-flexion models do not have. This implies the additional degrees of freedom granted by free Flexion are not constrained by the data alone, so we discarded flexion models for the final inference. Similar to H0LiCOW XII, we also find the kinematic weighting has little effect on the final inference of the Fermat potential. A corner plot of our posteriors is shown in \Cref{fig:corner_plot_final}, with final parameter estimates reported in \Cref{tab:final_parameters}.

\begin{table}[ht]
    \centering
    \begin{threeparttable}
    \caption{BIC+Kinematic-weighted posterior parameter estimates.}
    \renewcommand{\arraystretch}{1.5}
    \begin{tabular}{l|cc}
    \hline
    Parameter & This work (JWST) & H0LiCOW XII\tnote{‡} \ (HST) \\
    \hline
    $\Delta \tau_\mathrm{BC}$ & 0.469\lowup{0.009}{0.007} & 0.455\lowup{0.009}{0.010} \\
    $\Delta \tau_\mathrm{BA1}$ & 0.311\lowup{0.007}{0.005} & 0.298\lowup{0.007}{0.010} \\
    $\Delta \tau_\mathrm{BA2}$ & 0.330\lowup{0.008}{0.006} & 0.31\lowup{0.02}{0.04} \\
    $\Theta_\mathrm{E, PEMD}$ & 0.977\lowup{0.002}{0.002} & 0.93\lowup{0.02}{0.01} \\
    $\gamma_\mathrm{PEMD}$ & 1.86\lowup{0.01}{0.01} & 1.95\lowup{0.01}{0.02} \\
    $q_\mathrm{PEMD}$ & 0.774\lowup{0.007}{0.010} & 0.79\lowup{0.01}{0.01} \\
    $\phi_\mathrm{PEMD}$ & 30.3\lowup{0.5}{0.7} & 33.0\lowup{0.9}{0.8} \\
    $\gamma_\mathrm{ext}$ & 0.088\lowup{0.003}{0.003} & 0.112\lowup{0.004}{0.006} \\
    $\phi_\mathrm{ext}$ & 95\lowup{5}{5} & 84\lowup{2}{7} \\
    $\Theta_\mathrm{E, X}$ & 0.100\lowup{0.005}{0.005} & 0.0009\lowup{0.0007}{0.0010} \\
    $\Theta_\mathrm{E, G2}$ & 0.78\lowup{0.02}{0.02} & 0.93\lowup{0.06}{0.03}\tnote{‡‡} \\
    $\sigma_\mathrm{V}$ & 236\lowup{9}{20} & 232\lowup{7}{10} \\
    \hline
    \end{tabular}
    \tablefoot{Reported values are medians with errors corresponding to the 16th and 84th percentiles.}
    \begin{tablenotes}
            \footnotesize
            \item[‡] We note our weighting process may differ slightly for rounding discrepancies to values reported in H0LiCOW XII.
            \item[‡‡] The GLEE parameterization used an SIE model for G2, while we fit an SIS model due to degeneracies with the other perturbers.
        \end{tablenotes}
        \label{tab:final_parameters}
    \end{threeparttable}
\end{table}

\subsection{Satellite constraints}
\label{ssec:satellite_constraints}

The PEMD models in H0LiCOW XII found satellite X to have mass consistent with zero, which is indicative of poor constraints from the HST-based data. In addition, their models found an offset between the mass and light centroids for the main deflector of $0.02 - 0.03''$ ($\sim\!200$ pc, assuming a flat \text{$\Lambda \text{CDM}$} cosmology with $h = 0.7$ and $\Omega_{\rm m} = 0.3$). The influence of satellite X (lying to the northwest) was proposed as a partial potential explanation, as the centroid of the mass was found to be southeast of the light.

Our models place the Einstein radius of satellite X at $\theta_\text{E, X} = 0.100\lowup{0.005}{0.005}$, which is far closer to the theoretical value of $\theta_\text{E,X, init} \approx 0.15$ estimated in \Cref{ssec:mod_X}. Our models also find an offset between the mass and light centroids of the main deflector of $0.015 - 0.020''$,  which is smaller in scale compared to the HST models.

\subsection{External shear}
\label{ssec:external_shear}

The previous HST-based models were unable to explain the large shears required to model the system, even after including two large galaxy groups along the LOS (which we do not explicitly model here due to their minor impacts on the HST models but will consider in the following cosmographic analysis paper).
In this work, we have identified and robustly constrained the satellite galaxy X, which was only marginally detected in the analogous HST models. By constraining this satellite, we prevented the lens model from ``absorbing'' unmodeled mass distributions into a large external shear term. As a result, our models prefer a significantly smaller external shear, $\gamma_\mathrm{ext} = 0.088$, which is approximately $21\%$ lower than the values inferred by \cite{rusu2020}’s HST-based power-law models. This value is in close agreement with the observed shear estimates from \cite{wong2010}.

\subsection{Astrometry}
\label{ssec:astrometry}

\begin{figure}[ht]
	\centering
        Astrometry comparison: JWST vs HST \par\medskip
	\includegraphics[width=.5\textwidth]{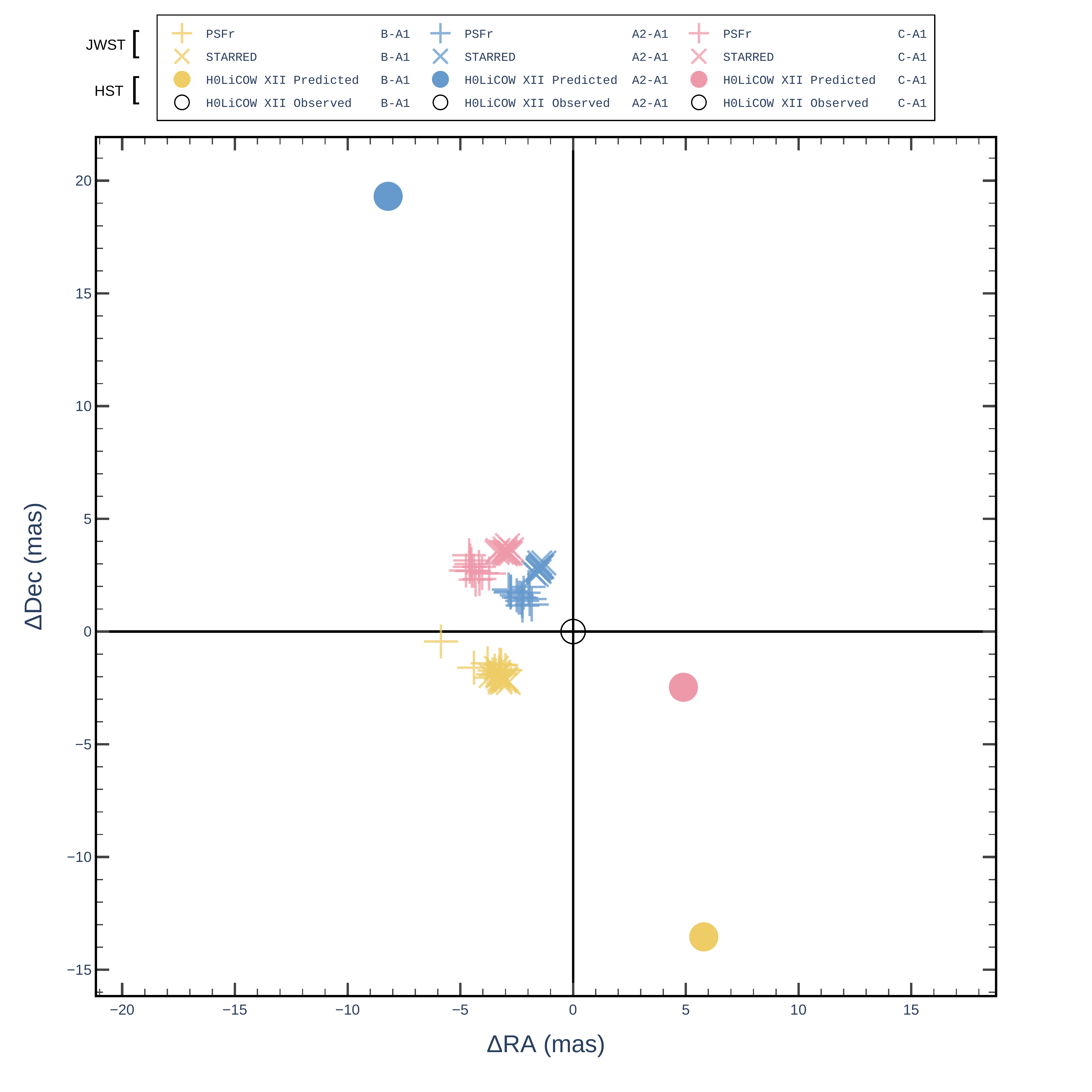}
	\caption{Relative image position discrepancies, centered based on the GLEE observed image positions. All distances are measured relative to the brightest image, A1. We note Lenstronomy requires the observed and predicted image positions to align. Image A2 also lies closest to the satellite, which is unconstrained in HST data, likely causing their largest discrepancy in image position. Our models demonstrate astrometry well within the requirements for time-delay cosmography \citep{birrer2019} and thus offer much tighter constraints than the previous HST-based GLEE models.}
	\label{fig:astrometry}
\end{figure}

In \Cref{fig:astrometry}, we compare the astrometric precision of our models to the H0LiCOW XII results. The expected relative positions of the images, centered at (0, 0), are derived from HST observations, with the brightest image (A1) serving as the reference point. Compared to these positions, our models achieve significantly smaller discrepancies than those constrained by the previous HST-based models. Specifically, we find deviations of $3.8 \, \text{mas}$ for the distance between images B and A1, $3.1 \, \text{mas}$ for A2 and A1, and $4.8 \, \text{mas}$ for C and A1. These are of the order of a tenth of a NIRCam pixel.

The overall relative astrometric uncertainty across all JWST models is $0.34 \, \text{mas}$, which we obtained by adding the uncertainties of individual images in quadrature. This is approximately 1/100 of a NIRCam pixel. Our estimated errors do not include potential residual systematics in the distortion correction of NIRCam, which should in any case very much be at the sub-pixel level, especially over a range of a few arcseconds covered by our target.

\section{Conclusion}
\label{sec:conclusion}

In this work, our aim was to extend the methods of time-delay cosmography to the first JWST imaging of quadruply imaged quasars in order to discover the benefits offered by the improved imaging while exploring potential systematic effects across telescopes. We modeled the system in the NIRCam F115W band, utilizing the spectroscopic data from \cite{sluse2019}, and applied the same wide-field data (external convergence) as \citealt{rusu2020} to enable a fair one-to-one comparison with the previous HST models. To accurately model the system, we implemented new tools to model the PSF of the system via \texttt{STARRED}. Our resulting Fermat potential difference is in agreement with the previous HST models at the $1\sigma$ level and well within expected uncertainties. The primary source of the discrepancy was the satellite, which was previously unconstrained in HST imaging. With JWST data, our models recovered a value in agreement with theoretical mass-to-light expectations. In summary, we found the following:

\begin{itemize}
    \item We achieved the first sub 2\% modeling uncertainty of any time-delay cosmography system, in agreement with previous HST models (\Cref{fig:FPs_comparison_HST_JWST}).

    \item We increased the accuracy of other key model parameters for this system, such as the power-law slope, Einstein radius, and axis ratio (\Cref{tab:final_parameters}).
    
    \item Modeling JWST's PSF with STARRED, we cut 71\% of the astrometric uncertainty compared to typical PSF-modeling techniques (\Cref{fig:astrometry}) and enabled a $17\%$ smaller scale of source reconstructions (\Cref{ssec:source_recons_supersampling}).
    
    \item We enhanced accuracy in the Fermat potential, driven by improved constraints on the satellite's mass (\Cref{ssec:satellite_constraints}). This work also provides a direct observational demonstration of how PEMD models with external shear can overcompensate for small-scale effects—below the resolution limit of the data—by adjusting the external shear \citep{etherington2023}.
    
\end{itemize}

We note this analysis was not done blinded to the model parameters in order to verify our handling of the PSF. However, the analysis was effectively blinded to $H_0$, as the system requires an updated sampling of the external convergence ($\kappa_\text{ext}$), which is a key ingredient in the final inference on $H_0$. In our next paper, we will make the complete inference on $H_0$ for this system, along with two other quadruply imaged quasars imaged with JWST.

For other future work, this analysis expands the sample of time-delay cosmography lenses to systems previously too faint to provide constraints on $H_0$. This is extremely complementary to the explosion of new lenses from future surveys such as the Rubin Observatory Legacy Survey of Space and Time (LSST) and Euclid, which will increase our sample of strong lens systems by two orders of magnitude \citep{collett2015}. Furthermore, developing methods to reliably constrain the PSF in fields with a limited number of field stars will be critical, as having six suitable reference stars in a field is unusual.

\begin{acknowledgements}
DW and TT acknowledge support by NSF through grants NSF-AST-1906976 and NSF-AST-1836016, and from the Moore Foundation through grant 8548.
Support for this work was provided by NASA through the NASA Hubble Fellowship grant HST-HF2-51492 awarded to AJS by the Space Telescope Science Institute, which is operated by the Association of Universities for Research in Astronomy, Inc., for NASA, under contract NAS5-26555. AJS also received support from NASA through the STScI grants HST-GO-16773 and JWST-GO-2974.
This work is also supported by JSPS KAKENHI Grant Numbers JP24K07089, JP24H00221. MS is partly supported by NASA through grant 80NSSC21K1294.

\end{acknowledgements}

\bibliographystyle{aa}
\bibliography{references}

\clearpage

\begin{appendix}

\FloatBarrier

\section{Parameter priors used}

\begin{strip}
	\centering
    \begin{threeparttable}
        \captionof{table}{Model parameter priors used, with some values rounded.}
	\begin{tabular}{lcllcc}
		\toprule
		Object & Component & Parameter & Distribution & Initial Position & Step Size (\boldmath$1\sigma$) \\
		\midrule

		\multirow{13}{*}{\shortstack[l]{Main Deflector (D) \\ $z=0.6575$}}
		& \multirow{4}{*}{\shortstack{Mass: \\ PEMD}} 
		&  $\theta_\text{E, PEMD}$      & $N(0.879, 0.133)$  & 0.879     & 0.133 \\
		& & $\gamma_\text{PEMD}$      & $U(1.5, 2.5)$  & 1.95     & 0.02 \\
		&                 & $(e_1, e_2)$         & $N(0, 0.2)$ & (-0.047, 0.107)   & 0.2 \\
		&                 & $(x, y)$     & $U(-3, 3)$     & (-0.031, 0.021)   & 0.003 \\
		\cmidrule(lr){3-6}
		& \multirow{4}{*}{\shortstack{Light: \\ Bulge Sérsic}} 
		& $R_\text{sersic}$      & $U(0.001, 5)$  & 0.296     & 0.01 \\
		&                 & $n_\text{sersic}$          & Fixed & 4   &  \\
		&                 & $(e_1, e_2)$         & $U(-0.2, 0.2)$ & (-0.048, 0.089)   & 0.1 \\
		&                 & $(x, y)$     & $U(-1, 1)$     & (-0.009, 0.009)  & 0.003 \\
		\cmidrule(lr){3-6}
		& \multirow{4}{*}{\shortstack{Light: \\ Disk Sérsic}} 
		& $R_\text{sersic}$      & $U(0.001, 5)$  & 1.752     & 0.05 \\
		&                 & $n_\text{sersic}$          & Fixed & 1   &  \\
		&                 & $(e_1, e_2)$         & $U(-0.5, 0.5)$ & (-0.164, 0.154)   & 0.1 \\
		&                 & $(x, y)$     & Joined with Bulge Sérsic     &   &  \\

		\midrule

		\multirow{3}{*}{\shortstack[l]{External Shear \\ $z=0.6575$}}
		& \multirow{3}{*}{\shortstack{Mass: \\ Shear}}
		& $\gamma_\text{ext, 1}$         & $U(-0.5, 0.5)$ & 0.109   & 0.1 \\
		&                 & $\gamma_\text{ext, 2}$         & $U(-0.5, 0.5)$ & 0.025    & 0.1 \\
		&                 & $(x, y)$     & Fixed     & 0   &  \\

		\midrule

		\multirow{7.5}{*}{\shortstack[l]{Satellite (X) \\ $z=0.6575$}}
		& \multirow{3}{*}{\shortstack{Mass: \\ SIS}}
		& $\theta_\text{E, X}$         & $U(0.0001, 0.25)$ & 0.15   & 0.01 \\
		&                 & $\gamma_\text{ext, 2}$         & $U(-0.5, 0.5)$ & 0.025    & 0.1 \\
		&                 & $(x, y)$     & Joined with Light      &    &  \\
		\cmidrule(lr){3-6}
		& \multirow{4}{*}{\shortstack{Light: \\ Sérsic}} 
		& $R_\text{sersic}$      & $U(0.0001, 0.1)$  & 0.044     & 0.01 \\
		&                 & $n_\text{sersic}$          & $U(1, 10)$ & 2   & 0.2 \\
		&                 & $(e_1, e_2)$         & Fixed & (0.015, 0.018)   &  \\
		&                 & $(x, y)$     & $U(\text{Initial} \pm 0.07)$      & (0.228, 2.046)   & 0.003 \\

		\midrule

		\multirow{6.5}{*}{\shortstack[l]{Galaxy G2 \\ $z=0.7450$}}
		& \multirow{2}{*}{\shortstack{Mass: \\ SIS}}
		& $\theta_\text{E, G2}$         & $N(0.622, 0.062)$ & 0.622   & 0.1 \\
		&                 & $(x, y)$     & Joined with Light      &    &  \\
		\cmidrule(lr){3-6}
		& \multirow{4}{*}{\shortstack{Light: \\ Sérsic}} 
		& $R_\text{sersic}$      & $U(0.001, 5)$  & 1     & 0.05 \\
		&                 & $n_\text{sersic}$          & $U(0.1, 2)$ & 0.5   & 0.1 \\
		&                 & $(e_1, e_2)$         & Fixed & (0, 0)   &  \\
		&                 & $(x, y)$     & Fixed      & (-3.998, -0.034)   &  \\

		\midrule

		\multirow{2}{*}{\shortstack[l]{Galaxy G3 \\ $z=0.6575$}}
		& \multirow{2}{*}{\shortstack{Mass: \\ SIS}}
		& $\theta_\text{E, G3}$         & $N(0.088, 0.048)$\tnote{†} &   &  \\
		&                 & $(x, y)$     & Fixed      & (2.823, 7.006)   &  \\

		\midrule

		\multirow{2}{*}{\shortstack[l]{Galaxy G7 \\ $z=0.6575$}}
		& \multirow{2}{*}{\shortstack{Mass: \\ SIS}}
		& $\theta_\text{E, G7}$         & $N(0.388, 0.030)$\tnote{†} &  &  \\
		&                 & $(x, y)$     & Fixed      &  (-4.578, -11.719)  &  \\

		\midrule

		\multirow{2}{*}{\shortstack[l]{Free Flexion \\ $z=0.6575$}}
		& \multirow{2}{*}{\shortstack{Mass: \\ Flexion}}
		& $(g_1, g_2, g_3, g_4)$         & $U(-0.1, 0.1)$ & 0  &  0.01\\
		&                 & $(x, y)$     & Fixed    &  (0, 0)  &  \\

		\midrule

		\multirow{4}{*}{\shortstack[l]{Quasar Images \\ $z=0.6575$}}
		& \multirow{5}{*}{\shortstack{Light: \\ Point Sources}} 
		& $\text{A1} (x,y)$      & $U(\text{Initial} \pm 0.3)$   & (-0.760, 0.958)     & 0.007 \\
		&                 & $\text{A2} (x,y)$          & $U(\text{Initial} \pm 0.3)$ & (-0.050, 1.076)   & 0.007 \\
		&                 & $\text{B} (x,y)$           & $U(\text{Initial} \pm 0.3)$ & (1.434, -0.302)   &  0.007\\
		&                 & $\text{C} (x,y)$           & $U(\text{Initial} \pm 0.3)$  & (-0.686, -0.579)    &  0.007\\

		\midrule

		\multirow{7.5}{*}{\shortstack[l]{Quasar Host Galaxy \\ $z=1.662$}}
		& \multirow{5}{*}{\shortstack{Light: \\ Sérsic}} 
		& $R_\text{sersic}$      & $U(0.04, 0.3)$  & 0.2     & 0.01 \\
		&                 & $n_\text{sersic}$          & $U(0.15, 4)$ & 1   & 0.25 \\
		&                 & $(e_1, e_2)$         & $U(-0.15, 0.15)$ & (0, 0)   &  0.03\\
		&                 & $(x, y)$     & $U(\text{Initial} \pm 1)$      & (-0.45, 0.1)   & 0.03 \\
		\cmidrule(lr){3-6}
		& \multirow{4}{*}{\shortstack{Light: \\ Shapelets}} 
		& $\beta$      & $U(0.031, 0.3)$  & 0.031     & 0.025 \\
		&                 & $n_\text{max}$          & Fixed & $\{18, 20, 22\}$   & 0.1 \\
		&                 & $(x, y)$     & Joined with Sérsic      &   &  \\
		
		\bottomrule
        \end{tabular}
        \tablefoot{Parameters grouped by components (Deflector, Sérsic 1, Shapelets) with their corresponding types (Mass or Light). The bounds are represented as uniform distributions, with initial guesses and sampling step sizes (\boldmath$\sigma$).}
        \begin{tablenotes}
            \footnotesize
            \item[†] While the Einstein radii are given priors, their actual values are scaled with G2. This ensures the constraints provided by their observed velocity dispersions are passed to the models, while also minimizing the degeneracies between their parameters.
        \end{tablenotes}
        \label{tab:components_parameters}
    \end{threeparttable}
\end{strip}

\FloatBarrier

\section{Verifying satellite X/galaxy G2 degeneracy}
\label{sec:XG2}

\begin{figure*}[ht]
	\centering
	\includegraphics[width=0.8\textwidth]{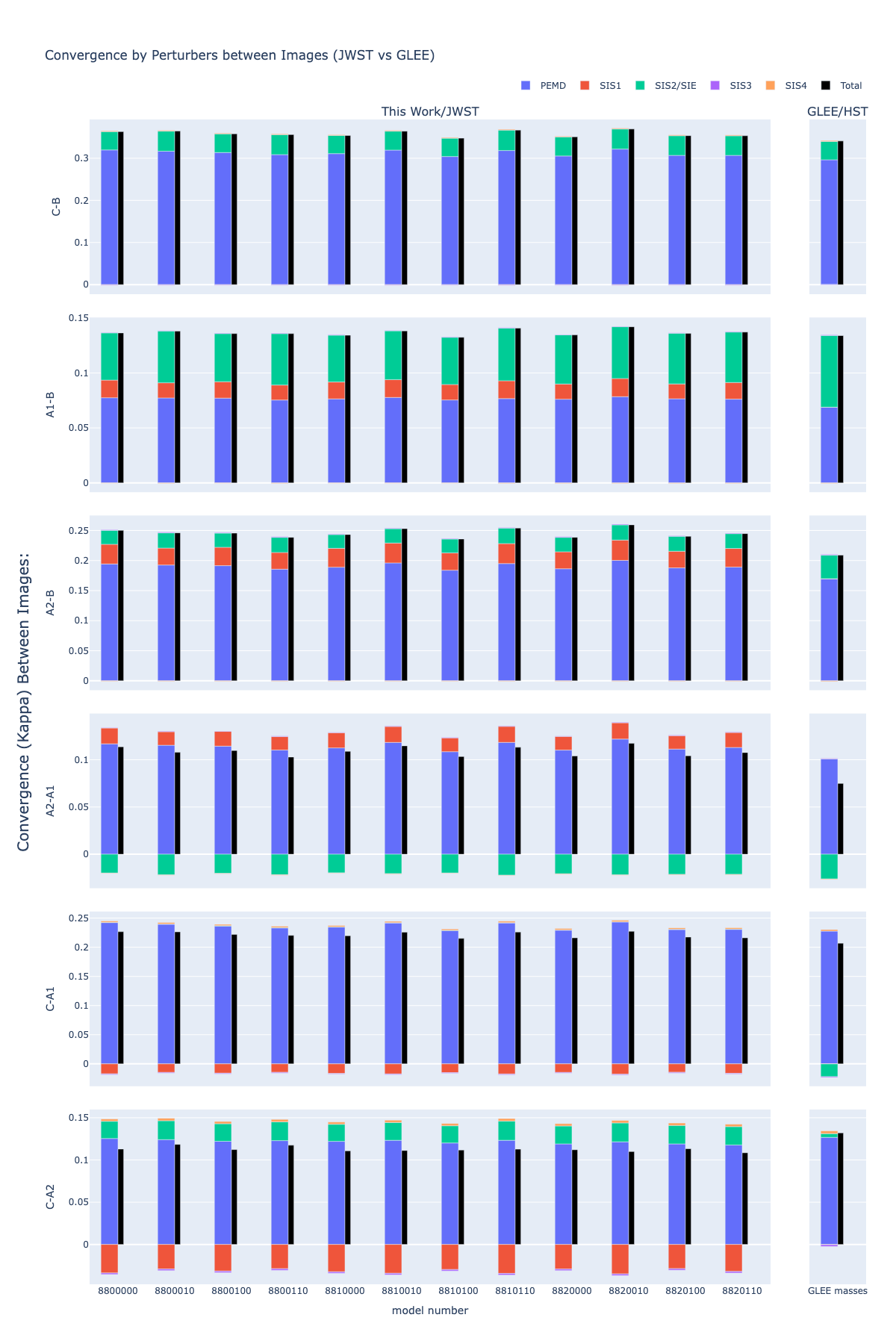}
	\caption{Differences in convergence between an image pair (vertical), broken down by each perturber. We compare each of our models (horizontal) with non-zero weight (left) to the HST results (right). Convergence difference from the main deflector is shown in blue, satellite X in red, nearby galaxy G2 in green, and the two farther perturbing galaxies G3 in purple and G7 in gold. We note that adding up the contributions from our models' spherical G2 and satellite X (green + red) almost perfectly matches the HST contributions from G2 alone (green). This supports our claim that the HST model's SIE profile for G2 is able to compensate for the missing satellite X, with the consequence of having an unphysical ellipticity and shifts in other model parameters.}
	\label{fig:convergence_comparison}
	\end{figure*}

We argue the HST models were able to replicate the effects of satellite X (which had a mass consistent with 0 in their power-law models) by allowing for an unphysical ellipticity in G2, when compared to its light profile. To test this degeneracy, we plotted the difference in convergence between different image pairs, broken down by each perturber, shown in \Cref{fig:convergence_comparison}. To compute the convergence for the HST results, we replaced the lens mass parameters in our models with those from their models, including the Einstein radius, slope, and ellipticity of the main deflector, Einstein radii of the perturbers, shear angle and magnitude, along with re-introducing the ellipticity of G2 resulting from the HST models. We reiterate that we did not allow G2 to be elliptical in our models, as we found the light profile of G2 did not match the ellipticities that the models required, indicative of an inherent degeneracy.

In the HST models, we find that the nearby galaxy G2 (green) is compensating for the convergence at each image position that should be coming from the satellite (red). This has a minimal impact on the final Fermat potential, as the total convergences match for the most important image pair (C-B, top). However, it helps explain the discrepancies in other parameters, such as the external shear and main deflector's Einstein radius and slope.

\FloatBarrier

\section{Additional systematic tests}
\label{subsec:appendix_additional_sys_tests}

\subsection{Polar exponential shapelet bases}

We tested representing our source complexity with polar exponential shapelet bases, which are theoretically expected to better encapsulate galaxy image information with fewer basis elements \citep{berge2019}. While our tests with polar exponential shapelets indeed yielded better BIC values, we found it came at the price of incorrect PSF reconstructions. This is likely caused by the increased information density near the origin of the shapelet model. This allowed for the extended source reconstruction to create point-source-like flux values at the center of the extended host's light distribution. This is problematic, as we need to accurately disentangle the light contributions from the point source and the extended source, otherwise our astrometry will be incorrect, biasing our results.

Therefore, despite the improved performance of polar exponential shapelets, we continue using the standard linear shapelets (e.g., H0LiCOW IX, \citealt{shajib2022}). We find their spatially uniform information density effectively prevents them from overfitting the center of the extended quasar host galaxy light.

\subsection{Two extended source sérsics}

All the source reconstructions from this system display a bulge-like component in the center of the host galaxy (see \Cref{fig:example_fit}). This bulge may have preference for a distinct Sérsic index from the extended disk structure of the host, as we observe in nearby galaxies. Therefore, instead of reconstructing the source with one Sérsic light profile modified by linear shapelets, we attempted fitting an additional Sérsic profile (maintaining the same scale of linear shapelets). We explored both leaving the indices free and fixing them to typical locally observed values (one de Vaucouleurs profile at $n_\text{sérsic} = 4$ and one exponential profile at $n_\text{sérsic} = 1$). We found the BIC slightly, but consistently, preferred the simpler one-Sérsic extended source model.

\subsection{Cutout size}

Although modeling larger cutouts aids in reconstructing the light profiles of the deflector and various perturbers, it comes at the cost of reduced weight on the extended source structure and drastic increases to computational time. To confirm the chosen mask sizes and positions ($2.1''$ and $2.2''$ radii) compose a sufficient representation of the lens system, we fit one model on a significantly larger cutout and found no statistically significant discrepancies.

\subsection{Fitting likely star-forming regions in host}
\label{ssec:SFR_regions}
In addition, we attempted to add additional shapelets to account for the residuals found south west of image C. We found the additional computational cost outweighed the negligible affect on parameter constraints.

\subsection{Scaling deflector Einstein radius with perturbers}
One possible modeling choice would be to scale the main deflector's Einstein radius with the rest of the galaxy-scale perturbers. We found it had negligible impact on the BIC and model results, but kept it free based on previous cosmography-grade lens models.

\subsection{Modeling the F150W band}
Test models were run on the F150W band to verify the precision and accuracy of the results when moving to larger wavelengths. No significant discrepancies were found in the inferred parameters, including the Fermat potential differences.

\subsection{Supersampling the extended source light}
As supersampling the data is computationally intensive, regions that are supersampled are typically confined to the centers of the point sources. To test for potential improvements on parameters such as $\gamma_\text{PEMD}$, we ran tests that also supersampled a large portion of the extended light of the host. We found the computational trade-off outweighed the marginal gain on parameter precision.

\subsection{Freeing lens light Sérsic profiles}
We found leaving the Sérsic indices free for the main deflector's light profiles led to an additional degeneracy with their resulting radii. This slowed model convergence, while having no noticeable impact on the resulting Fermat potential differences.

\FloatBarrier

\section{Model weights}
The model weights used for the final inference are shown in \Cref{tab:res_mods}. The models including flexion are weighted separately, as they are not used in the final inference.

\begin{table*}
	\centering
	\caption{Final models sorted by total weight (kinematics weight $\times$ modeling/BIC weight).}
	\label{tab:res_mods}
	\begin{tabular}{cccccccc}
		\toprule
		Main Deflector & Source (Shapelet Order) & Perturbers & PSF Type & Mask (asec) & $\Delta$BIC & Total Weight \\
		\bottomrule
		\multicolumn{7}{c}{No Free Flexion} \\
		\midrule
            SPEMD + 2 SERSIC & 18 $n_\mathrm{max}$ + SERSIC & X+G2+G3+G7 & STARRED & 2.2 & 38 & 1.000 \\
            SPEMD + 2 SERSIC & 20 $n_\mathrm{max}$ + SERSIC & X+G2+G3+G7 & STARRED & 2.2 & 96 & 0.483 \\
            SPEMD + 2 SERSIC & 20 $n_\mathrm{max}$ + SERSIC & X+G2+G3+G7 & STARRED & 2.1 & 190 & 0.094 \\
            SPEMD + 2 SERSIC & 18 $n_\mathrm{max}$ + SERSIC & X+G2+G3+G7 & STARRED & 2.1 & 195 & 0.084 \\
            SPEMD + 2 SERSIC & 22 $n_\mathrm{max}$ + SERSIC & X+G2+G3+G7 & STARRED & 2.2 & 303 & 0.005 \\
            SPEMD + 2 SERSIC & 22 $n_\mathrm{max}$ + SERSIC & X+G2+G3+G7 & STARRED & 2.1 & 410 & 0.000 \\
            SPEMD + 2 SERSIC & 18 $n_\mathrm{max}$ + SERSIC & X+G2+G3+G7 & PSFr & 2.2 & 1968 & 0.000 \\
            SPEMD + 2 SERSIC & 18 $n_\mathrm{max}$ + SERSIC & X+G2+G3+G7 & PSFr & 2.1 & 2014 & 0.000 \\
            SPEMD + 2 SERSIC & 20 $n_\mathrm{max}$ + SERSIC & X+G2+G3+G7 & PSFr & 2.2 & 2538 & 0.000 \\
            SPEMD + 2 SERSIC & 20 $n_\mathrm{max}$ + SERSIC & X+G2+G3+G7 & PSFr & 2.1 & 2164 & 0.000 \\
            SPEMD + 2 SERSIC & 22 $n_\mathrm{max}$ + SERSIC & X+G2+G3+G7 & PSFr & 2.1 & 2502 & 0.000 \\
            SPEMD + 2 SERSIC & 22 $n_\mathrm{max}$ + SERSIC & X+G2+G3+G7 & PSFr & 2.2 & 2379 & 0.000 \\
		\hline
		\multicolumn{7}{c}{Free Flexion} \\
		\hline
            SPEMD + 2 SERSIC & 18 $n_\mathrm{max}$ + SERSIC & X+G2+G3+G7+Flexion & STARRED & 2.1 & 0 & 0.000 \\
            SPEMD + 2 SERSIC & 18 $n_\mathrm{max}$ + SERSIC & X+G2+G3+G7+Flexion & STARRED & 2.2 & 27 & 0.000 \\
            SPEMD + 2 SERSIC & 20 $n_\mathrm{max}$ + SERSIC & X+G2+G3+G7+Flexion & STARRED & 2.1 & 63 & 0.000 \\
            SPEMD + 2 SERSIC & 20 $n_\mathrm{max}$ + SERSIC & X+G2+G3+G7+Flexion & STARRED & 2.2 & 66 & 0.000 \\
            SPEMD + 2 SERSIC & 22 $n_\mathrm{max}$ + SERSIC & X+G2+G3+G7+Flexion & STARRED & 2.2 & 226 & 0.000 \\
            SPEMD + 2 SERSIC & 22 $n_\mathrm{max}$ + SERSIC & X+G2+G3+G7+Flexion & STARRED & 2.1 & 257 & 0.000 \\
            SPEMD + 2 SERSIC & 18 $n_\mathrm{max}$ + SERSIC & X+G2+G3+G7+Flexion & PSFr & 2.1 & 1702 & 0.000 \\
            SPEMD + 2 SERSIC & 18 $n_\mathrm{max}$ + SERSIC & X+G2+G3+G7+Flexion & PSFr & 2.2 & 1778 & 0.000 \\
            SPEMD + 2 SERSIC & 20 $n_\mathrm{max}$ + SERSIC & X+G2+G3+G7+Flexion & PSFr & 2.2 & 1852 & 0.000 \\
            SPEMD + 2 SERSIC & 20 $n_\mathrm{max}$ + SERSIC & X+G2+G3+G7+Flexion & PSFr & 2.1 & 2007 & 0.000 \\
            SPEMD + 2 SERSIC & 22 $n_\mathrm{max}$ + SERSIC & X+G2+G3+G7+Flexion & PSFr & 2.2 & 2091 & 0.000 \\
            SPEMD + 2 SERSIC & 22 $n_\mathrm{max}$ + SERSIC & X+G2+G3+G7+Flexion & PSFr & 2.1 & 2218 & 0.000 \\
		\bottomrule
	\end{tabular}
    \tablefoot{The weighting is separated between the flexion and non-flexion model families. The $\Delta \text{BIC}$ values are calculated relative to the best performing (lowest-BIC) model. For more information, see \Cref{subsec:comb_sys}.}
\end{table*}

\FloatBarrier

\section{Impact of PSF noise estimation}
\label{ssec:psf_noise_estimation}

It is relatively challenging to estimate the additional uncertainty on the data caused by inaccuracies of the PSF model. However, as long as the residuals of the PSF are sufficiently amplified (without incorrectly amplifying the uncertainty of the extended source structure), this should have minimal impact on model results as shown in previous analyses (e.g., H0LiCOW XII).

\begin{figure}[htpb!]
    \vspace{-0.54cm}
	\centering
	\includegraphics[width=0.355\textwidth]{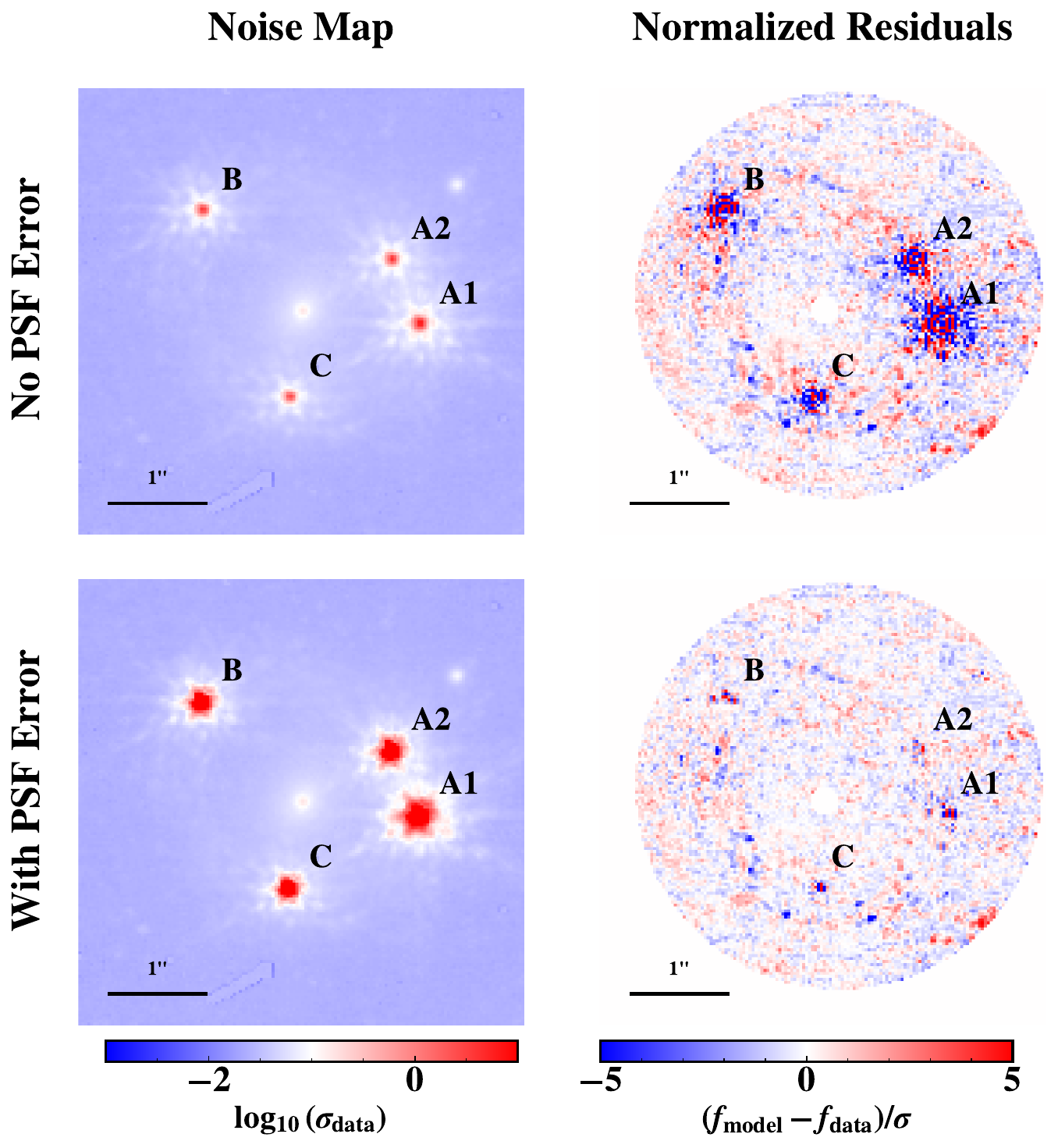}
	\caption{
    Differences in normalized residuals when correcting for uncertainty in the PSF (this example used the STARRED PSF model). Without this correction, the fitting of the point sources drives the fit while providing little information. Top left: Pixel uncertainty derived from the imaging data alone. Top right: Model-normalized residuals without PSF uncertainty correction. Bottom left: Pixel uncertainty including additional PSF uncertainty, estimated by fitting stars in the field (\cref{subsec:mod_init_psf}). Bottom right: Model-normalized residuals with PSF uncertainty correction. Residuals are more evenly distributed, encouraging the models to accurately describe the ring structure.}
	\label{fig:psf_error_residuals}
\end{figure}

\begin{figure*}[htpb!]
	\centering
	\includegraphics[width=\textwidth]{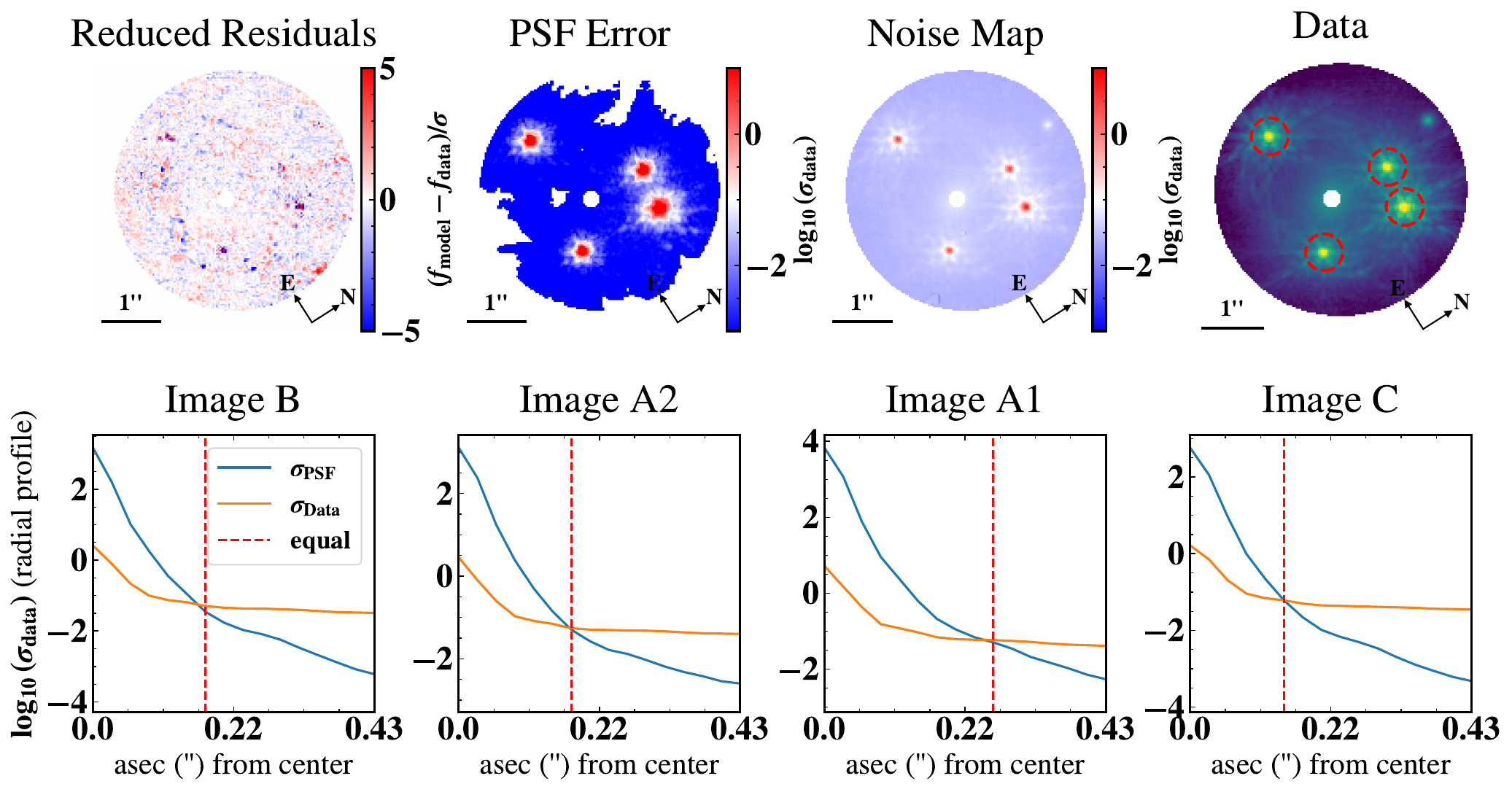}
	\caption{
    Radial breakdown of image uncertainty contributions between the noise directly in the data and the additional PSF uncertainty term, showing the PSF contribution does not down-weight the ring structure of the extended source. Top left: normalized residual map, same as \cref{fig:psf_error_residuals}. Top middle left: error contributions from the PSF uncertainty alone, and top middle right shows only the uncertainty due to the data alone. These are combined to form the final map, shown in \cref{fig:psf_error_residuals}. Top right: the imaging data used for modeling, with circles indicating the largest radial extent of the images where the PSF uncertainty equals the noise uncertainty. Shown on the bottom are the radial uncertainty contributions from the imaging data vs the estimated PSF error. We find the maximum extent at which the PSF error dominates the uncertainty is at a radius less than 11 pixels ($\approx0.338''$), shown in the top right figure.}
	\label{fig:psf_error_radial_profile}
\end{figure*}

The residuals of our models--after removing this additional PSF uncertainty in the data--showed no discernible trends that would raise concern (\Cref{fig:psf_error_residuals}). We also note minimal overlap between regions with amplified noise and the extended host structure, which proved a non-trivial task with JWST's PSF (\Cref{fig:psf_error_radial_profile}). This is crucial to avoid down weighting the ring structure of the extended source, which provides the tightest constraints on the power-law slope.
\FloatBarrier
\section{Tests on source reconstruction supersampling}
\label{ssec:source_recons_supersampling}
Given that this represents the most detailed source reconstruction ($n_\text{max}={18, 20, 22}$) among time-delay cosmography systems modeled with Lenstronomy, care must be taken to avoid supersampling beyond a numerically feasible limit. If the number of shapelets (or similarly, the shapelet scale) pass the threshold of information supported by the data, the model will be attempting to resolve spatial scales smaller than the data's pixel scale would constrain. This would lead to a nonphysical solution in the source light reconstruction, as the surface brightness of the source would no longer be conserved.

As we use linear shapelets for the parameterization of our source reconstruction, the minimum source scale reconstructed\footnote{Technically, the ``true" source scale depends on the spatially dependent source plane magnification. However, this magnification always increases the minimum allowed source scale. In this case, the analysis that follows applies even more strongly.} is given by
\begin{equation}
	L_\text{min} = \frac{\beta}{\sqrt{n_\text{max}+1}}.
\end{equation}
The shapelet scales reconstructed by our PSFr models were at $\beta_\text{PSFr} = 0.089\lowup{0.005}{0.010}$, while the STARRED models reconstructed at $\beta_\text{STARRED} = 0.074\lowup{0.004}{0.006}$ (STARRED model scales were on average $17\%$ smaller than their PSFr counterparts). Ideally, this reconstruction scale should not be smaller than the spatial scale of our pixels, as this would imply that we are reconstructing at scales not constrained by the data. In other words, we require
\begin{equation}
	L_\text{min} \ge \delta_\text{pix},
\end{equation}
with our pixels in the drizzled F115W band having a spatial resolution of $\delta_\text{pix} \sim 0.0307\arcsec/\text{pix}$.

However, when supersampling the source, this form changes to reflect the increased source resolution assumed to be constrained by the supersampled data. For a supersampling factor of 3, the spatial scales in the image plane that constrain the source ray tracing are 3 times smaller than the true pixel scale of the data, yielding the new constraint
\begin{equation}
	L_\text{min} \ge \delta_\text{pix}/f_\text{ss},
\end{equation}
with our models taking a supersampling factor of $f_\text{ss} = 3$. This yields a minimum reconstruction scale allowed by the data of $L_\text{min} = 0.0103$, while the smallest value sampled by any of our models was $L_\text{min, model} = 0.0126$, still above the threshold for numerical stability.

\FloatBarrier

\section{STARRED versus PSFr parameter comparison}
\label{sec:starred_vs_psfr_corner}

\begin{figure*}[htbp!]
	\centering
	\includegraphics[width=\textwidth]{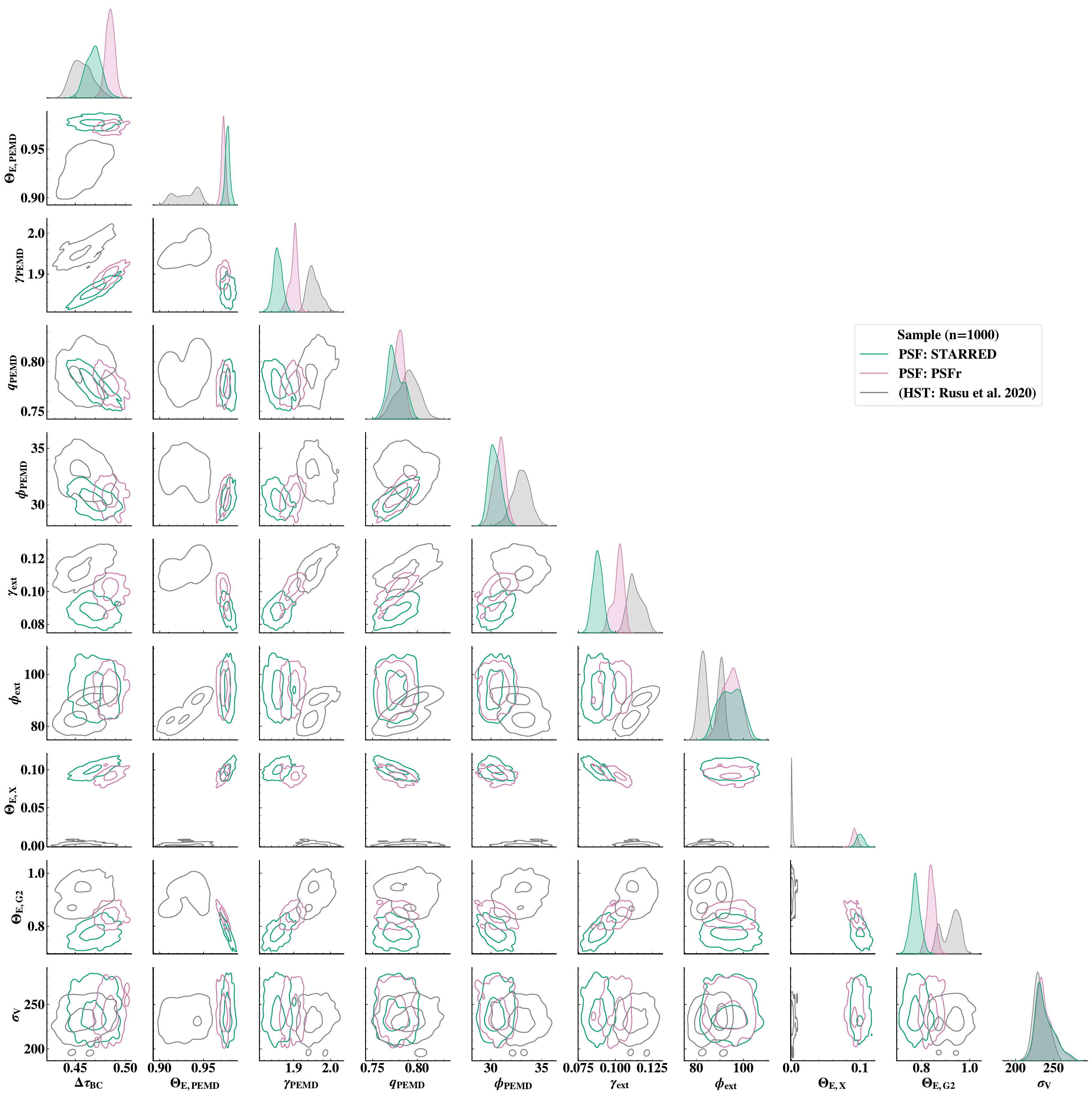}
	\caption{Same as \cref{fig:corner_plot_final}, except comparing weighted PSFr models to weighted STARRED models. Both groups receive BIC weights independently for this comparison.}
	\label{fig:starred_psfr_corner}
\end{figure*}

\cref{fig:starred_psfr_corner} shows cornerplots of the non-flexion models (as flexion was discarded for our results), comparing models using PSFr vs STARRED to generate their initial PSF models. The two samples are assigned weights separately. STARRED had eight models with non-zero weights while PSFr had three. Despite both PSFs being updated during fitting, the disagreement in initial PSF leads to a more than 1-$\sigma$ disagreement in parameters such as the power-law slope, demonstrating the significance of accurate PSF modeling.

\end{appendix}

\end{document}